%% file: bao.tex
\documentclass[notitlepage,nofootinbib,twocolumn,amsmath]{revtex4-1}

\usepackage{graphicx}
\usepackage{amsmath}
\usepackage{amssymb}
\usepackage{amsfonts}
\usepackage{aas_macros}
\usepackage{bm}
\usepackage{color}
\usepackage{hyperref}
\usepackage[caption = false]{subfig}

\input{defs.tex}

\allowdisplaybreaks[1]

\begin{document}
\title{Phenomenology of BAO evolution from Lagrangian to Eulerian Space}
\author{Tobias Baldauf}
\email{t.baldauf@damtp.cam.ac.uk}
\affiliation{Centre for Theoretical Cosmology, DAMTP, University of Cambridge, CB3 0WA, UK}
\author{Vincent Desjacques}
\email{dvince@physics.technion.ac.il}
\affiliation{Physics department, Technion, Haifa 3200003, Israel,}
\affiliation{D\'epartement de Physique Th\'eorique
and 
Center for Astroparticle Physics, 
Universit\'e de Gen\`eve, CH-1211 Gen\`eve, Switzerland}


\begin{abstract}
The baryon acoustic oscillation (BAO) feature provides an important distance scale for the measurement of the expansion history of the Universe. 
Theoretical models of the BAO in the distribution of biased tracers of the large scale structure usually rely on an initially linear BAO.
With aid of N-body simulations, we demonstrate that the BAO in the initial (Lagrangian) halo 2-point function is significantly sharper than in the
linear matter distribution, in agreement with peak theory. Using this approach, we delineate the scale-dependence induced by the higher-derivative 
and velocity bias before assessing how much of the initial BAO enhancement survives until the collapse epoch. Finally, we discuss the extent to 
which the velocity or gravity bias, which is also imprinted in the displacement field of halos, affects the contrast of the BAO obtained with a 
reconstruction.
\end{abstract}

\maketitle

\section{Introduction}

The Baryon Acoustic Oscillation (BAO) feature is a standard distance scale imprinted into the cosmic density fluctuations. 
In the correlation function it manifests itself as a distinct peak at $r_\text{BAO}\approx 108 \hMpc$, while it leads to a series of oscillations 
in the Fourier space power spectrum \cite{peebles/yu:1970,sunyaev/zeldovich:1970}.
This feature is also present in the clustering statistics of tracers of Large-Scale Structure (LSS), such as galaxies and their host halos
\cite{cole/2df:2005,eisenstein/sdss:2005}.
Predictions for its shape and position are instrumental for the measurement of this distance scale and the inference of the expansion history 
of the Universe.

High mass halos have been shown to form out of the maxima of the underlying density field \cite{Ludlow:2011pk}. This approximation is reasonably
accurate down to a mass of a few $M_*(z)$, where $M_*(z)$ is the characteristic mass of fluctuations virializing into halos at a given redshift
$z$. Besides this phenomenological evidence for halo formation sites being associated with the maxima of the initial Gaussian random field - the
so-called peak-patches \cite{bond/myers:1996a,bond/myers:1996b,bond/myers:1996c}, the clustering of maxima pioneered by \cite{Bardeen:1985tr} 
shows a rich phenomenology which goes beyond that of local Lagrangian bias models \cite{fry/gaztanaga:1993}. 
In particular, the dependence on the curvature $\partial^2\delta_\text{L}$ of the linear density field through the peak constraint sharpens the linear 
theory BAO contrast in Lagrangian space \cite{Desjacques:2008jj}. 
Moreover, owing to the correlation between linear velocities $\sim \partial_i^{-1}\delta_\text{L}$ and gradients $\partial_i\delta_\text{L}$ 
\cite{Bardeen:1985tr}, the displacement field of Lagrangian peak-patch exhibits a scale-dependent bias which affects
also the shape of the final BAO \cite{Desjacques:2010gz}. Thus far however, numerical studies of these effects and comparison with theory are still 
limited. Using peak theory and a phase space conservation argument, Ref.\cite{Desjacques:2010gz} derived a prediction for the halo 2-point correlation 
within the Zel'dovich approximation, which they compared to the $z=0$ halo correlation function extracted from a large N-body simulation. 
Nevertheless, their peak model did not properly take into account the cloud-in-could problem \cite{epstein:1983,appel/jones:1990,bond/cole/etal:1991}. 
Furthermore, Refs.\cite{elia/ludlow/porciani:2012,Baldauf:2015ve} measured in the initial conditions of their simulations a velocity bias $c_v(k)$ 
consistent with the peak prediction for $M\gtrsim M_*$, but these measurements were limited to Fourier space and did not address the time-evolution 
of the BAO feature. 

In this paper, we will revisit these issues and illustrate how peak theory explains the shape of the BAO in the 2-point correlation function of 
both Lagrangian peak-patch halos (or, simply, proto-halos) and evolved Eulerian halos. Unlike \cite{Desjacques:2010gz}, we will rely on the excursion 
set peak approach \cite{paranjape/sheth:2012,Paranjape:2012jt}, which incorporates (at least part of) the cloud-in-cloud constraint, to describe the 
Lagrangian peak-patches. We will compare our predictions with measurements of the Lagrangian and Eulerian halo 2-point correlation across the BAO 
feature. Finally, we will use the peak approach to ascertain how much the BAO can be sharpened by a reconstruction, and whether the peak velocity
bias matters for reconstruction. Hence, we extend the discussion of \cite{noh/white/padmanabhan:2009} to biased tracers which are not described by a
simple local Lagrangian biasing scheme (see \cite{biasreview} for areview on bias).

Our paper is organized as follows. In Sec.~\ref{sec:BAOLag}, we begin with a comparison between model predictions and simulations in Lagrangian space.
In Sec.~\ref{sec:BAOEul}, we discuss the time-evolution of the 2-point peak correlation from a phase space perspective using the Zel'dovich 
approximation, and compare our prediction at $z=0$ with the 2-point correlation of simulated dark matter halos. We discuss the relevance of the 
peak approach for BAO reconstruction in Sec.~\ref{sec:BAOrec}. We conclude in Sec.~\ref{sec:conclusion}. Technical details can be found in an
Appendix (\ref{app:za}). In all illustrations, a $\Lambda$CDM cosmology compatible with cosmic microwave background (CMB) and LSS measurements was 
assumed.

\section{Baryon Oscillation in Lagrangian space}
\label{sec:BAOLag}

\subsection{BAO smoothing in the initial field}

The correlation function of peaks in a Gaussian random field can be calculated by filtering the field at a Lagrangian scale $R$ 
and correlating points with vanishing first derivative and negative definite Hessian \cite{Bardeen:1985tr}.
For definiteness, we will employ a Gaussian filter in this study, although one could adopt different filters for different 
variables (e.g. \cite{Paranjape:2012jt}).
In Lagrangian space, the leading order density and momentum statistics of peaks can be computed from the linear biasing relations
\cite{Desjacques:2008jj}
\begin{align}
\delta_\text{h}(\vec k) &= c_1(k,z_c)\, \delta_\text{L}(\vec k) \label{eq:biasdpk} \\
\vec j_\text{h}(\vec k) &= c_v(k)\, \vec j_\text{L}(\vec k) \;, \label{eq:biasvpk} 
\end{align}
where $\delta_\text{L}(\vk)$ and $\vec j_\text{L}(\vk)$ are the linear matter density and momentum fields extrapolated to the collapse 
redshift $z_c$ (we shall omit their explicit redshift dependence for conciseness), and
\begin{align}
c_1(k,z_c) &\equiv \Big(b_{10} + b_{01} k^2\Big) W_R(k) \label{eq:c1L} \\
c_v(k) &\equiv \Big(1-R_v^2 k^2\Big)W_R(k) \label{eq:bv}
\end{align}
are the first-order, Lagrangian peak bias function and the linear peak velocity bias , respectively. Note that $c_1(k,z_c)$ 
depends on redshift, whereas $c_v(k)$ does not. Furthermore, $b_{10}$ is the usual linear, Lagrangian density bias $b_1^L$. 
To avoid clutter, we will omit the superscript ``$L$'' on all the Lagrangian quantities, but use the superscript ``$E$'' for the
evolved, Eulerian quantities. 
The appearance of a linear velocity bias is a selection effect reflecting the peak constraint \cite{Bardeen:1985tr,Desjacques:2009kt}. 
As a result, the gravitational force acting on halo centers is also biased 
(see, e.g., \cite{Desjacques:2010gz,Baldauf:2015ve,biagetti/desjacques/etal:2016,biasreview}).

The $k^2$ term in the Fourier space bias function $c_1(k,z_c)$ corresponds to the appearance of a higher-derivative operator 
$\bm{\nabla}^2\delta_R$ -- where $\delta_R$ is the linear density field smoothed on scale $R$ -- in the perturbative peak bias 
expansion. This term leads to an enhancement in the halo matter power spectrum 
in the mildly non-linear regime before the window function cuts of the power on very small scales corresponding to inter-halo scales. 
In \cite{Baldauf:2015ve} (hereafter BDS), it was shown that this model provides a good description of the initial halo-matter density 
and momentum statistics in $N$-body simulations and, in particular, with the measurements of $b_{01}$
(see also \cite{elia/ludlow/porciani:2012,modi/castorina/seljak:2016}).
The fits to the halo-matter power spectrum typically require $R=0.5 R_\text{TH}$, where $R_\text{TH}=(3M/4\pi \bar\rho)^{1/3}$ is the 
radius of the top hat filter containing the halo mass.
The leading order halo-halo and halo-matter Lagrangian power spectra predicted by this model are given by \cite{Desjacques:2009kt}
\begin{align}
P_\text{hh} &\supset c_1^2(k,z_c)\,P_\text{L}(k) \label{eq:phhL} \\
P_\text{hm} &= c_1(k,z_c)\, P_\text{L}(k) \;, \label{eq:phmL}
\end{align}
where, again, our convention implies that the linear power spectrum $P_\text{L}(k)$ is evaluated at the redshift $z_c$ of halo collapse.
For sake of conciseness, we have omitted both the higher order and the shot-noise terms in the halo-halo power spectrum. 
In the correlation function, the higher-derivative operator $\bm{\nabla}^2\delta_R$ generates first-order contributions 
$-2 b_{10}b_{01}\bm{\nabla}^2\xi_R$ and $(b_{01})^2\bm{\nabla}^4\xi_R$, where $\xi_R$ is the linear, smoothed correlation function. 
These contributions are particularly sensitive to features in the correlation function, such as the BAO peak for instance 
\cite{Desjacques:2008jj}. It should also be noted that this linear, higher-derivative bias is different from higher order biases 
such as $b_2$ for instance, which induce scale dependence through loop corrections.

\begin{table}
\caption{Best-fit values for the Lagrangian, linear halo bias parameters as determined in BDS. The halo mass is in unit of $10^{13}\hmsun$.
The values quoted in parenthesis are those used for the predictions shown in Fig.\ref{fig:finalbao}. Note that $R_v^2-b_{01}<0$, which 
corresponds to an enhanced BAO feature at late time.}
\vspace{1mm}
\begin{center}
\begin{tabular}{c|c|c|c|c}
\hline
 & Mass & $b_{10}$ & $b_{01}$ & $R_v$ \\
\hline\hline
bin I & ~0.78~  & 6.48$\times 10^{-2}$ & 4.89 ~ (7.26)  & 1.99 ~ (2.45) \\
\hline
bin II & ~2.3~ & 0.312 & 10.3 ~ (15.7) & 2.98 ~ (3.30) \\
\hline
bin III & ~6.9~ & 0.818 & 26.4 ~ (34.0) & 3.97 ~ (4.43) \\
\hline
bin IV & ~20~ & 1.69 & 51.4 ~ (67.3) & 5.45 ~ (5.78) \\
\hline
\end{tabular}
\end{center}
\label{table1}
\end{table}

\begin{figure*}
\subfloat{\includegraphics[width=0.49\textwidth]{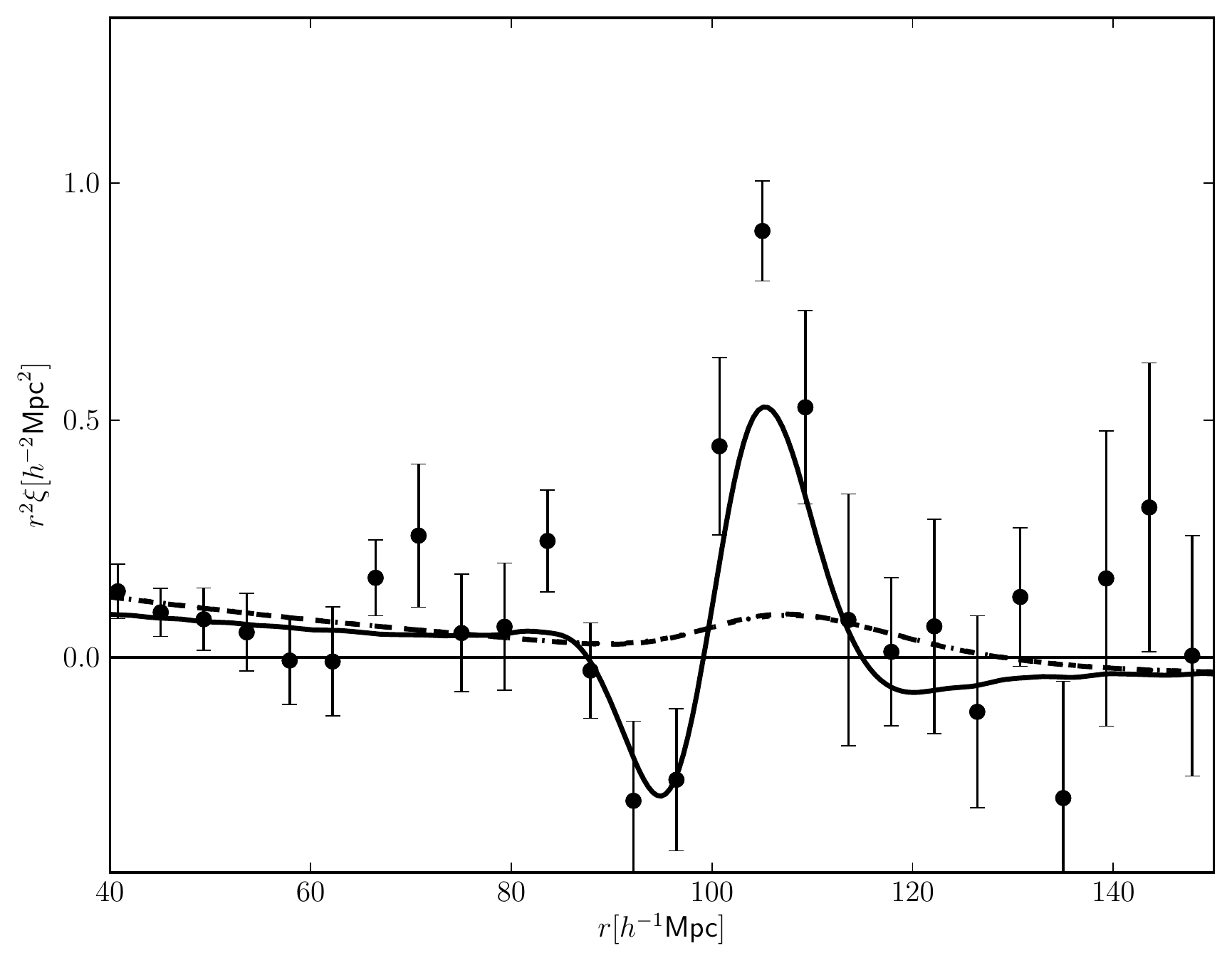}}
\subfloat{\includegraphics[width=0.49\textwidth]{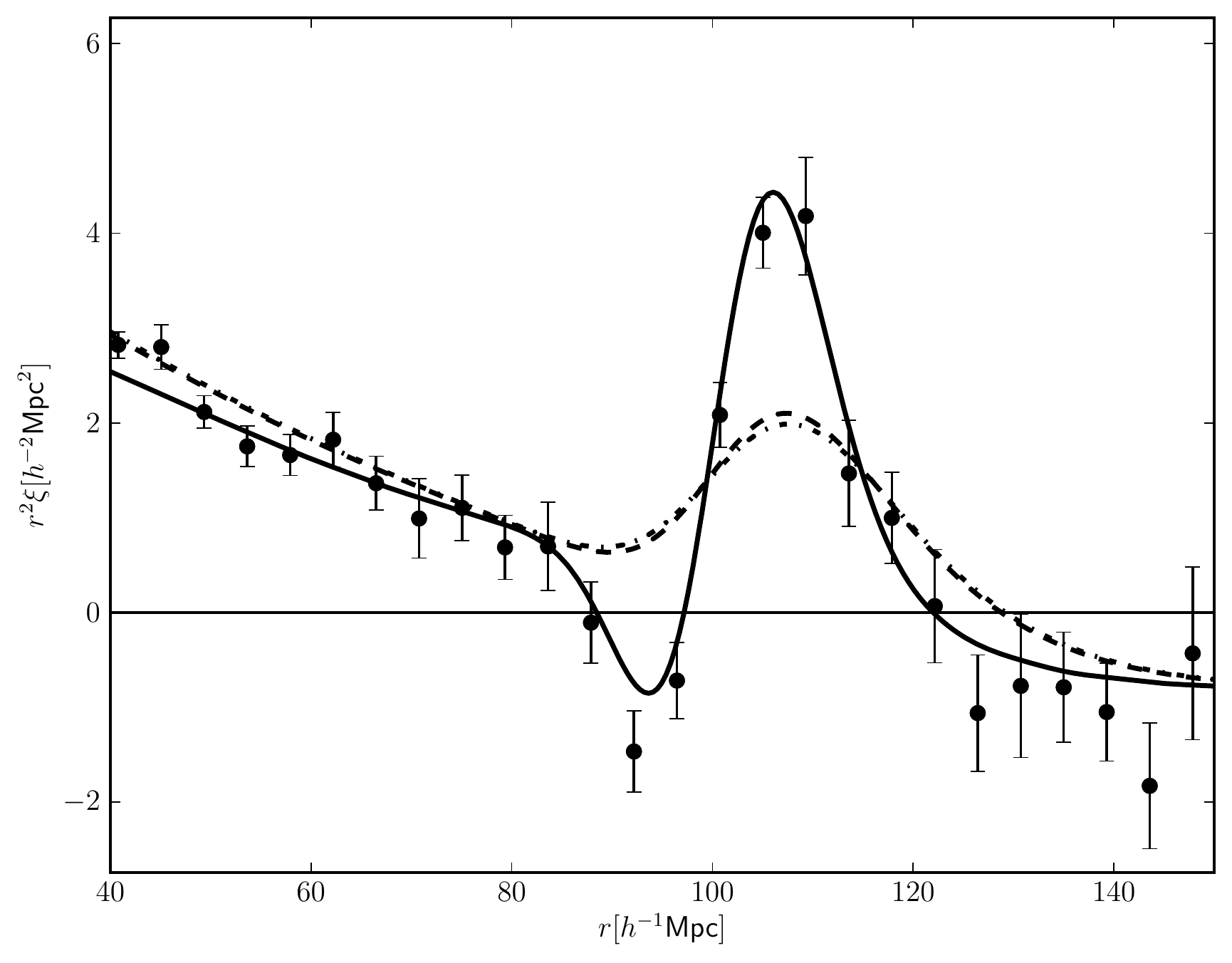}} \\
\subfloat{\includegraphics[width=0.49\textwidth]{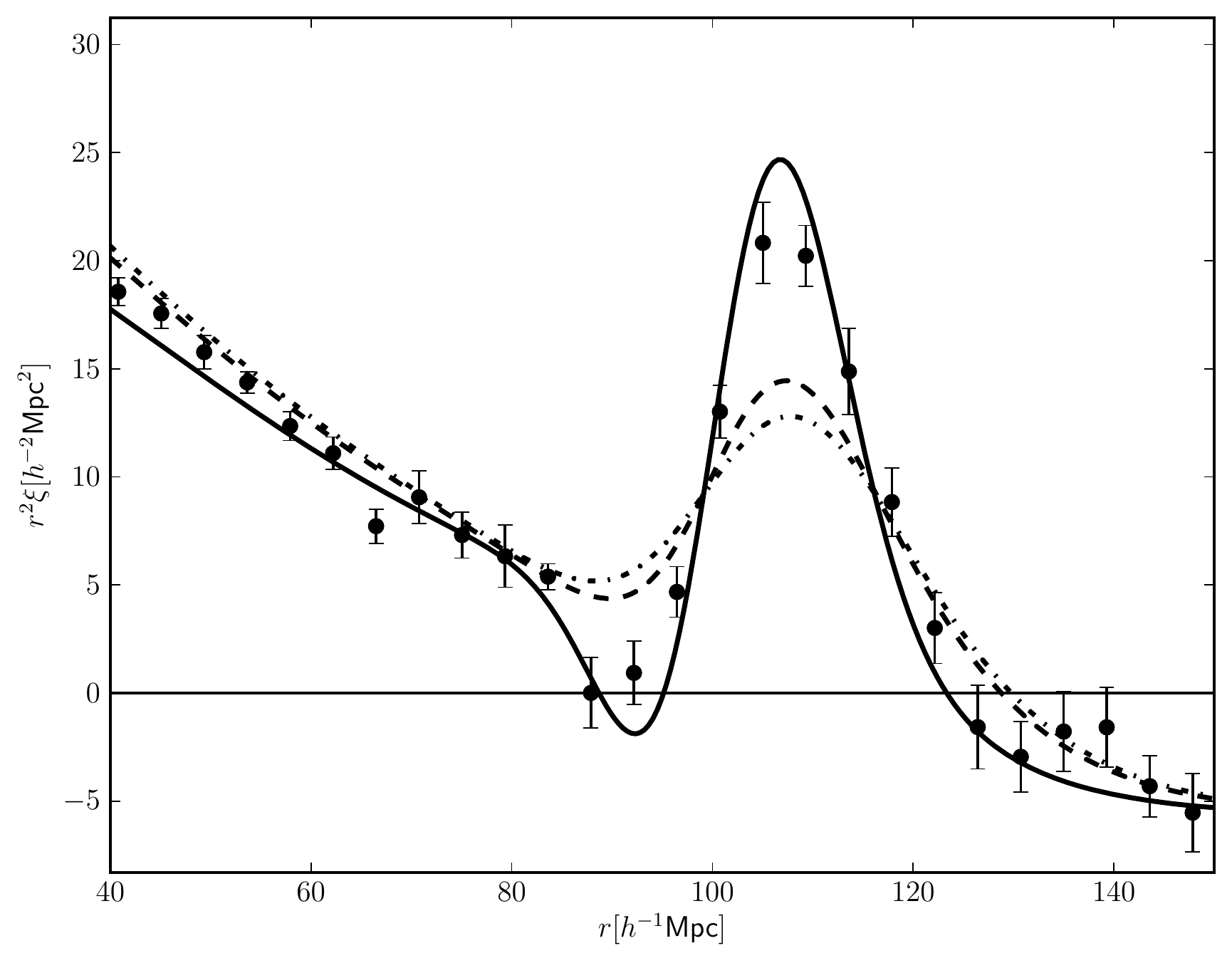}}
\subfloat{\includegraphics[width=0.49\textwidth]{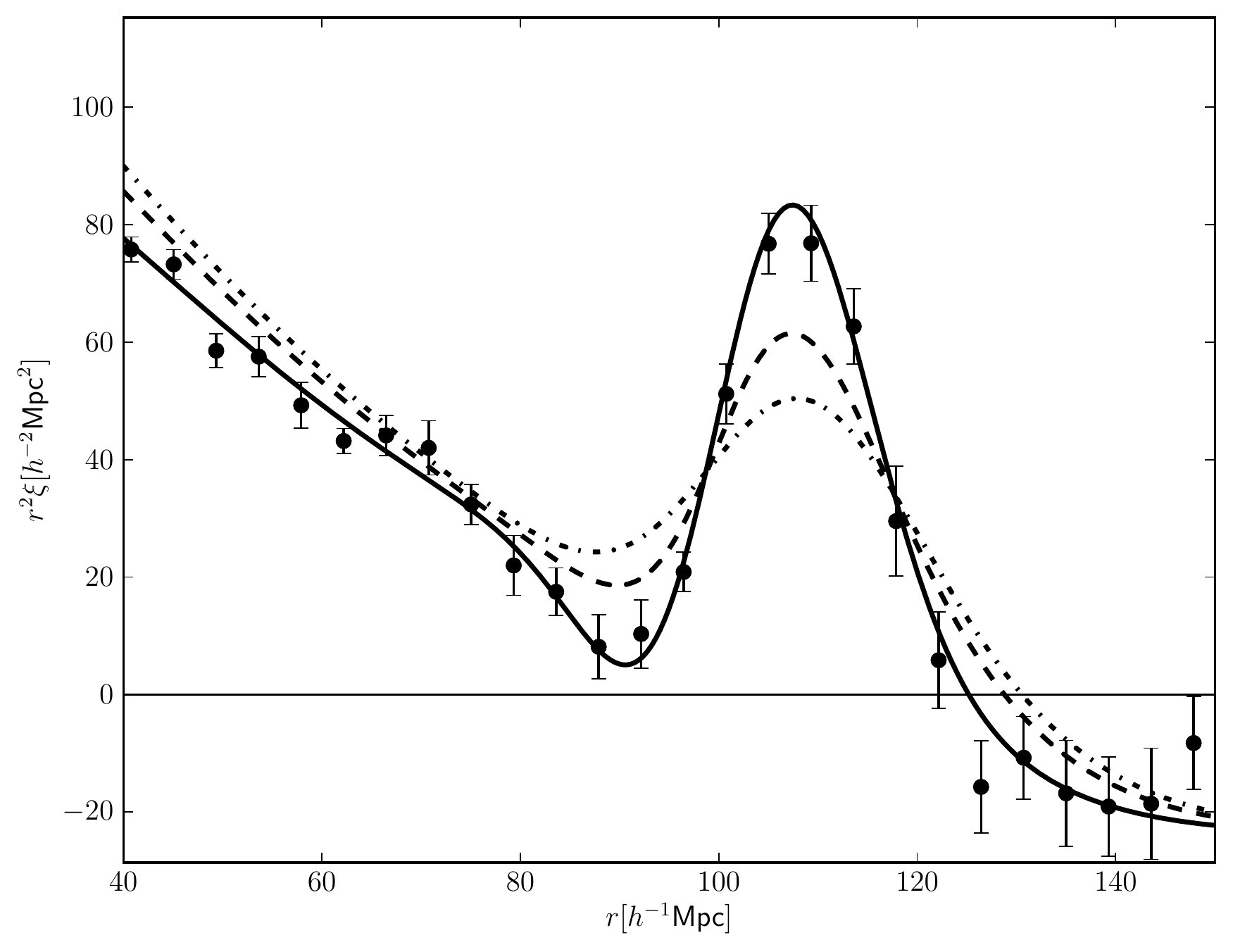}}
\caption{Enhancement of the BAO in the initial conditions for the halo bins I -- IV from top left to bottom right. 
Note that, for some of the mass bins, the correlation is negative even at separations smaller than the BAO scale.
The thick solid curve represents the peak model prediction, while the dashed and dashed-dotted curves indicate the 
local bias and the smoothed, linearly biased correlation function, respectively.
The linear correlation function is not a good model, and the smoothed correlation function is even worse.
}
\label{fig:initialbao}
\end{figure*}

In Fig.~\ref{fig:initialbao}, we compare the prediction Eq.(\ref{eq:phhL}) obtained with the peak bias parameters determined in BDS 
-- the best-fit values of $b_{10}$, $b_{01}$ and $R_v$ are given in Table \ref{table1} -- to the Lagrangian BAO peak of the same halo 
sample. For illustration, we also show the linearly biased linear theory correlation function $b_{10}^2 \xi_\text{L}$, the linearly 
biased smoothed linear theory correlation function $b_{10}^2 \xi_R$. 
We clearly see that the peak model is in excellent agreement with the measured BAO peak, whereas both the linear theory and smoothed 
linear theory predictions fail at reproducing its shape. This Lagrangian BAO peak provides the initial conditions for a dynamical 
understanding of the Eulerian BAO peak as in \cite{Desjacques:2010gz} or, alternatively, in CLPT \cite{Carlson:2013co} or iPT 
\cite{Matsubara:2011ck}.

\subsection{Halo displacement dispersion}

Moments of the displacement from Lagrangian to Eulerian space are a key ingredient of any Lagrangian Perturbation Theory (LPT, see
\cite{Zeldovich:1970,buchert:1989,moutarde/alimi/etal:1991,hivon/bouchet/etal:1995,Matsubara:2007wj}).
In the Zel'dovich approximation \cite{Zeldovich:1970}, the Fourier modes of the displacement field $\vec\Psi$,
\be
\label{eq:linearDisp}
\vec\Psi(\vk) = i \vec L^{(1)}(\vk) \delta_\text{L}(\vk) \qquad \mbox{with} \qquad \vec L^{(1)}(\vk)=\frac{\kvh}{k} \;,
\ee
are proportional to those of the (linear) velocity $\vec v$, i.e. $\Psi_i(\vk)=v_i(\vk)/\mathcal{H}f$ \cite{Bouchet:1994xp}.
Therefore, in order to facilitate the comparison, we will rescale all velocities to correspond to redshift zero displacements.
In peak theory, the 1st order LPT displacement shares the functional form of Eq.(\ref{eq:linearDisp}), with $\vec L^{(1)}(\vk)$ 
replaced by
\be
\vec L^{(1)}_\text{pk}(\vk) = \frac{\kvh}{k} c_v(k)\;.
\ee
Consequently, this implies that the linear (3-dimensional) velocity dispersion of halos is given by
\cite{Bardeen:1985tr}
\be\label{eq:sigmavpk}
\la\vec\Psi^\top\vec\Psi|{\rm pk}\ra
=\sigma_{-1}^2-\frac{\sigma_0^4}{\sigma_{1}^2} 
\equiv \sigma_{v,\text{pk}}^2\;,
\ee
where
\be
\sigma_i^2=\int_{\vk} k^{2i} W_R^2(k) \, P_\text{L}(k) \;,
\ee
Here, $\sigma_{v,\text{m}}\equiv\sigma_{-1}$ and $\sigma_{v,\text{pk}}$ are the linear velocity dispersion of the dark matter and the peaks,
respectively, and $\int_{\vk}$ is a shorthand for the momentum integral $\int\! d^3k/(2\pi)^3$.
The first term in the right-hand side differs from the linear dark matter in the presence of the explicit smoothing scale, whereas the
second term always contributes negatively. Therefore, we expect that, in the Zel'dovich approximation, the displacement dispersion
of halos is reduced relative to the dark matter. 

\begin{figure}
\includegraphics[width=0.5\textwidth]{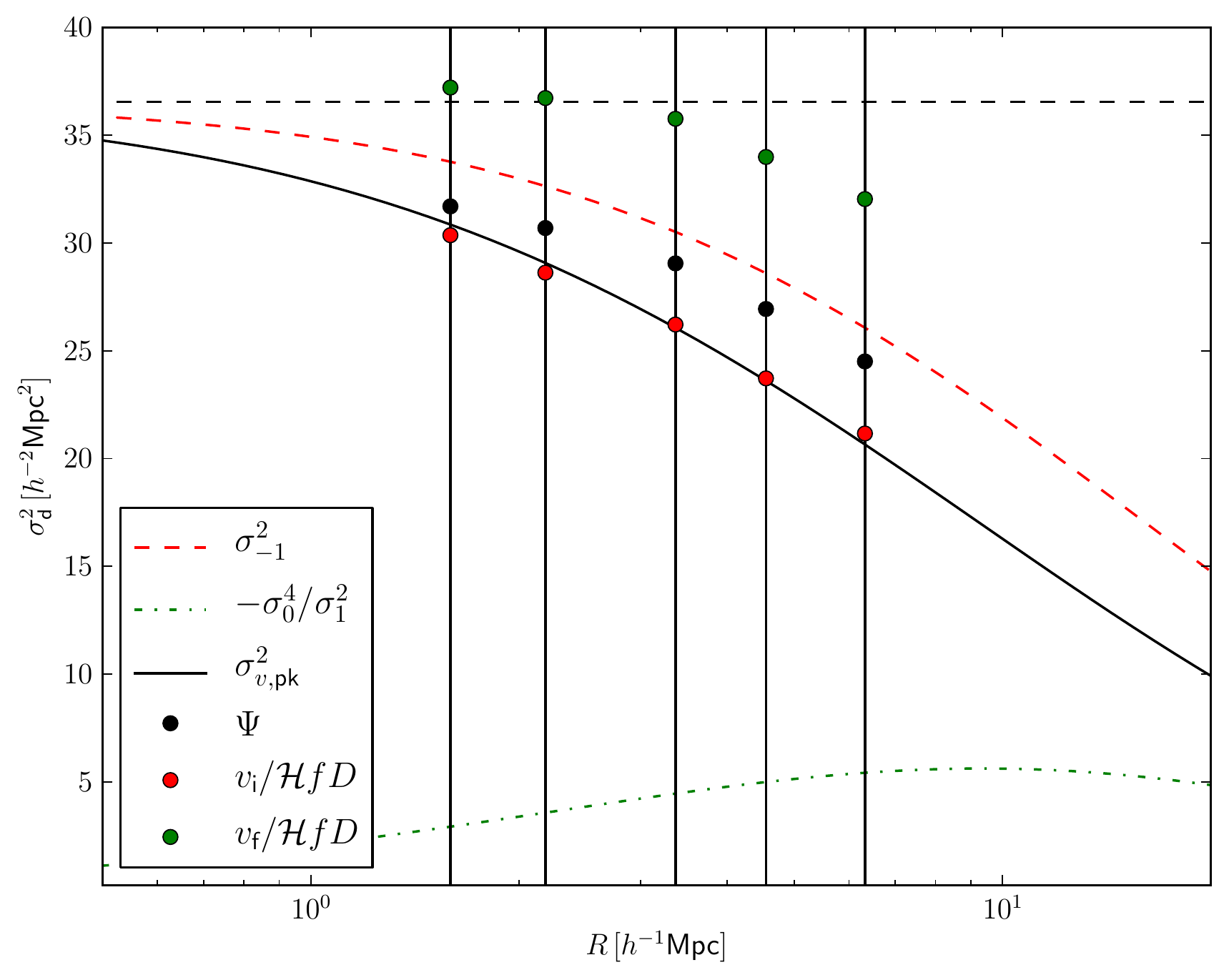}
\caption{The 3-dimensional displacement and velocity dispersions of halos. For convenience we have rescaled the velocities by 
$\mathcal {H} f$ such that they have units of displacement. 
The horizontal line represents the unsmoothed, linear dark matter velocity dispersion.
The red dashed curve shows the effect of smoothing, while the green dashed curve indicate the absolute value of the correction 
from the peak constraint. 
The Lagrangian proto-halo velocities are in very good agreement with the peak model. The total halo displacement dispersion is 
suppressed relative to linear theory matter displacement, but exceeds the Lagrangian rms.} 
\label{fig:halveldisp}
\end{figure}

In Fig.\ref{fig:halveldisp}, the data points indicate the measurements of the initial and final {\it number-weighted} velocity dispersion 
together with the {\it number-weighted} displacement dispersions for the halos extracted from our simulations. 
For convenience, we have rescaled velocities by $\mathcal{H}f D$ to have units of Zel'dovich displacement to redshift zero. 
The various curves represent the various terms in Eq.(\ref{eq:sigmavpk}). The horizontal line is the linear velocity dispersion of the 
unsmoothed dark matter velocity field.
Results are shown as a function of the Gaussian smoothing scale $R$, which is related to the Lagrangian size $R_M=(3M/4\pi\bar\rho_m)^{1/3}$ 
of the halo according to $R\approx 0.5 R_M$. 
The initial halo velocity dispersion is very well described by the peak model prediction. Whereas, for small halos, the peak velocity 
dispersion and our measurements converge towards the unsmoothed linear theory displacement dispersion, they are significantly suppressed 
for massive halos. 
The dominant part of this suppression arises from the explicit smoothing scale with additional suppression coming from the peak constraint 
encoded by the second term in Eq.~\eqref{eq:sigmavpk}.
If the Zel'dovich displacement provided a full description of the halo motions, the measured displacement field should be in perfect 
agreement with the rescaled initial velocities. This is clearly not the case: the displacement dispersion exceeds the initial velocity 
dispersion for all masses, but is still significantly suppressed with respect to unsmoothed linear theory. 
The departure from the initial velocity dispersion arises from acceleration with respect to the ballistic Zel'dovich displacement. 
The origin for the latter may be twofold: firstly, there are undoubtedly higher order displacement fields in Lagrangian Perturbation theory 
which contribute to the full displacement \cite{Bouchet:1994xp,Scoccimarro:1997gr,Matsubara:2007wj}. Secondly, we speculate that the 
shrinking of the effective halo radius once the collapse sets in, as predicted in simple collapse models \cite{Gunn:1972sv}, may also 
impact velocities immediately, while the total displacement would be affected only later.
Notwithstanding, we will here and henceforth stick to the Zel'dovich approximation, and treat $R$ as the Lagrangian filter radius, so that
it remains constant throughout halo collapse.

\section{Baryon Oscillation in Eulerian space}
\label{sec:BAOEul}

Pairwise motions induced by inhomogeneities in the mass distribution distort the Lagrangian correlation of halo centers. 
Assuming that the latter are test particles which locally flow with the dark matter, the observed 2-point correlation function of 
virialized halos is modelled as a convolution of the Lagrangian (initial) one, $\xhh(r,z_i)$, with the displacement field of initial 
density peaks. This can be accomplished upon, e.g., writing down this convolution explicitly \cite{Desjacques:2010gz} using phase
space distributions (see \cite{davis/peebles:1977,bharadwaj:1996} for earlier work) or within the so-called Integrated Perturbation
Theory (iPT) \cite{Matsubara:2011ck}. 

\subsection{Preliminary considerations}
\label{sec:keff}

\begin{figure}
\includegraphics[width=0.5\textwidth]{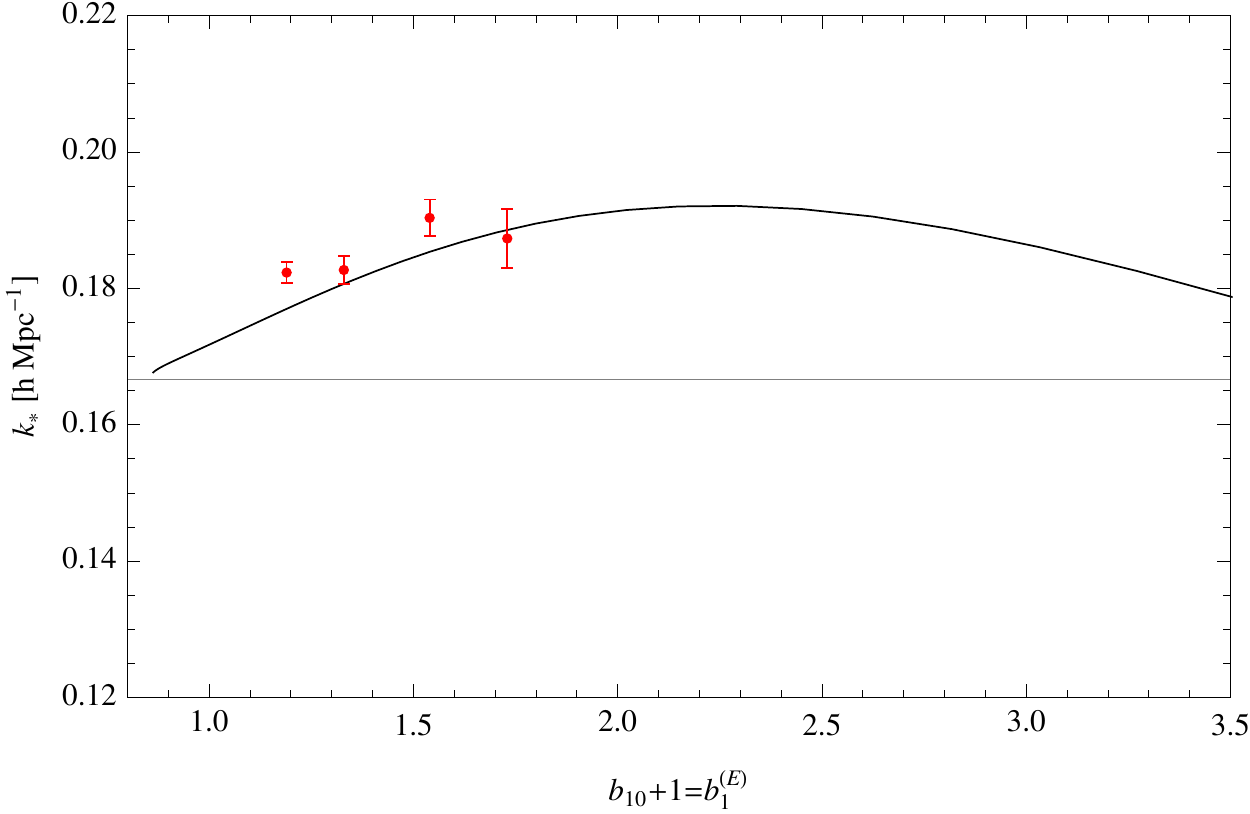}
\caption{Effective BAO smoothing scale $k_*$ as a function of the (Eulerian) linear halo bias. 
We compare the measurements of \cite{Prada:2014bra} to the prediction Eq.~\eqref{eq:effbaosmoothing} based on the peak model, which 
includes the initial peak spatial and velocity bias as well as the reduced peak velocity dispersion. Both the data and our prediction 
exhibit a larger softening wavenumber $k_*$ (i.e. a reduced BAO broadening). The magnitude and the scaling with linear bias agree 
reasonably well.}
\label{fig:kstar}
\end{figure}

Displacements with wavelength larger than the BAO width ($\sim 10\hMpc$) but smaller than the BAO scale ($\sim 100\hMpc$) tend to 
broaden the Lagrangian BAO. For dark matter, the leading-order contribution in  the Zel'dovich approximation for the BAO oscillations 
reads
\footnote{Due to the equivalence principle, the long wavelength motions have to cancel for the broadband. Therefore, this equation 
should be regarded as an approximation which captures the dominant effects for the real space BAO feature \cite{Baldauf:2015xfa}.}
\be
P_\text{mm}^E=\exp\!\left( -\frac{1}{3} k^2 \sigma_{v,\text{m}}^2 \right) P_\text{L}(k) \; ,
\ee
where $\sigma_{v,\text{m}}$ is the dispersion of the linear, 3-dimensional matter velocity field.
For halos, the BAO smoothing differs due to the presence of the Lagrangian BAO enhancement, the scale dependence of the velocity bias 
and the difference between the halo and dark matter displacement dispersions \cite{Desjacques:2009kt,Desjacques:2010gz}. We have
\begin{align}
P_\text{hh}^E &\supset \Big[c_v(k) + c_1(k,z_c) \Big]^2 e^{-\frac{1}{3}k^2\sigma_{v,\text{pk}}^2}\, P_\text{L}(k) 
\label{eq:phhtree} \\ 
P_\text{hm}^E &\supset \Big[c_v(k) + c_1(k,z_c) \Big] 
e^{-\frac{1}{6} k^2\sigma_{v,\text{pk}}^2-\frac{1}{6} k^2\sigma_{v,\text{m}}^2}\, P_\text{L}(k) \nonumber \;.
\end{align}
The effective smoothing induced by the BAO broadening can be associated to a scale upon collecting all the $k^2$ correction terms and 
factorizing out the large scale linear bias. In the halo-matter power spectrum, this yields
\be\label{eq:effbaosmoothing}
P_\text{hm}^E(k)\approx \big(1+b_{10}\big)\Bigg[1-\frac12 \bigg(\frac{k}{k_*}\bigg)^2\Bigg] 
e^{-\frac{1}{6} k^2\sigma_{v,\text{m}}^2} P_\text{L}(k) \;,
\ee
where 
\be\label{eq:keffbaosmoothing}
k_* \equiv \Bigg(\frac{1}{3}\sigma_{v,\text{pk}}^2+R^2+2\frac{R_v^2-b_{01}}{1+b_{10}}\Bigg)^{-1/2} \;.
\ee
In Fig.~\ref{fig:kstar}, we compare this prediction to the detailed measurements of the BAO smoothing of \cite{Prada:2014bra} extracted 
from $N$-body simulations. These authors performed a series of high resolution simulations with and without BAO wiggles in the initial 
power spectrum, but identical random seeds. They were thus able to cancel cosmic variance among simulations with the same seeds, and 
perform a detailed measurement of the BAO damping.
Our prediction assumes the best-fit values given in Table \ref{table1}. It agrees reasonably well with the data of \cite{Prada:2014bra}
for their range of halo mass. 
We conclude that the combination of smoothing window and peak selection effects in the initial density and displacement statistics can 
indeed account for the reduced smoothing of the halo with respect to the dark matter BAO peak. 

\subsection{Peak correlations in the Zel'dovich approximation}

In general, Eulerian matter clustering statistics can be expressed in terms of Lagrangian correlators using phase space conservation 
(see, e.g., \cite{bharadwaj:1996}). Let $\vec\Psi_{12}=\vec\Psi_2-\vec\Psi_1$ be the relative displacement between two tracers.
In the Zel'dovich approximation and under the peak constraint, the 2-point correlation function can be formulated as the convolution 
\cite{Desjacques:2010gz} (see Appendix \ref{app:za} for a brief derivation of this relation)
\begin{widetext}
\be
\label{eq:xpkzexact}
\bnh^2\left[1+\xpk^E(r,z_c)\right]=
\int_{\vk}\int \!d^3q\,e^{i\vk\cdot(\vr-\vq)}
\int\! d^N\! y_1 d^N\! y_2\,\npk(\bm{y}_1)\npk(\bm{y}_2)
p_2(\bm{y}_1,\bm{y}_2;\vq,z_i)\,
e^{-\frac{1}{2}\sigma_{v,\text{m}}^2\vk^\top\vcc\,\vk-i\sigma_{_{v,\text{m}}}\vk^\top\Delta\vec\Psi}\;,
\ee
\end{widetext}
where $\sigma_{v,\text{m}}$ is the rms dispersion of the dark matter at the collapse redshift $z_c$ and $p_2$ is the joint distribution 
of the peak state vectors $\bm{y}_1$ and $\bm{y}_2$ in Lagrangian space.
For the peak constraint considered here, $\bm{y}=(\nu,\eta_i,\zeta_A)$  where (in the notation of \cite{Bardeen:1985tr}) $\nu$ is the
peak height, $\eta_i$ are the three components of the normalized derivative and $\zeta_A$ are the six independent entries of the Hessian
$\zeta_{ij}$ (i.e. $N=10$).
Furthermore, $\vr$, $\vq$ are the initial and final pair separation vector, respectively, whereas 
$\Delta\vec\Psi(\vq)=\la\vec\Psi_{12}|\bm{y}\ra$ is the relative, average displacement conditioned to the value of the vector $\bm{y}$
and $\vcc(\vq)=\la (\vec\Psi_{12}-\Delta\vec\Psi)\otimes(\vec\Psi_{12}-\Delta\vec\Psi)|\bm{y}\ra$ is the corresponding covariance matrix.
Note that, in the conditioning $\lvert\bm{y}\rangle$, the exact nature of the tracer is still unspecified (except for the fact that
it depends on $\bm{y}$).
The covariance matrix can be expressed as $\vcc=2(\vcc_{\vec 0}-\vcc_{\vr})$, where $\vcc_{\vec 0}$ and $\vcc_{\vr}$ are $3\times 3$ block 
matrices describing the correlations at zero lag and at finite separation, respectively.
Note that $\vcc$ becomes singular when $\vr=\vec 0$. This explicit dependence on the relative displacement $\vec\Psi_{12}$ which, at 
leading order, is directly proportional to the deformation tensor $\partial_i\partial_j\Phi$, shows that only second- or higher-order 
derivatives of the potential lead to observable effect in the peak correlation function.
Consequently, Eq.~\ref{eq:xpkzexact} is invariant under the Extended Galilean transformations (GI) \cite{rosen:1972}, i.e. the uniform 
but time-dependent boosts 
\be
\vx'=\vx+\vx_0(t) \;, \qquad t'=t
\ee
where the vector field $\vx_o(t)$ is an arbitrary function of time. 
This ensures that the effect of very long wavelength perturbations vanishes in the equal-time, 2-point peak correlation \ref{eq:xpkzexact}
\cite{scoccimarro/frieman:1996,kehagias/riotto:2013,peloso/pietroni:2013,kehagias/norena/etal:2014,valageas:2014,Baldauf:2015xfa}. 
We will come back to this point shortly.

Eq.\ref{eq:xpkzexact} is a direct consequence of Liouville's theorem, i.e. the phase space conservation of the tracers, which is trivially 
ensured by the one-to-one mapping between the Lagrangian peak -- patches and the virialized halos. 
In practice, the numerical evaluation of \ref{eq:xpkzexact} is fairly tedious owing to the high dimensionality of the integral. Hence, 
we shall now briefly discuss a couple of tractable approximations.

\subsubsection{Perturbative expansion in Fourier space}
\label{sub:kspace}

The first option consists in expanding the integrand in small correlation functions and perform the integral $\int d^3q$ over the Lagrangian
separation, so that $\xpk^E(r,z)$ is explicitly written as a Fourier transform. This was the approach adopted in \cite{Desjacques:2010gz}. 
In practice, the displacement field is decomposed onto a basis aligned with the Lagrangian separation vector $\vr=\vq_2-\vq_1$, where $\vq_i$
are the initial positions of the peak patch.  
As a result, the covariance matrix $\vcc$ of the relative peak displacement $\vec\Psi_{12}$ becomes block-diagonal. 
The calculation of the tree-level contribution, along with a sketch of the derivation of the 1-loop mode-coupling, is presented in Appendix 
\ref{app:za_kspace}. After some manipulations, we eventually arrive at
\begin{widetext}
\be
\label{eq:xpkz1loop}
\xpk^E(r,z_c) = \int_{\vk}\,\bigg\{e^{-\frac{1}{3}k^2\sigma_{v,\text{pk}}^2} 
\Bigl[c_1^E(k,z_c)\Big]^2 P_\text{L}(k)+P_\text{1-loop}(k,z_c)\bigg\}e^{i\vk\cdot\vr} + \mathcal{O}(\text{2-loop}) \;.
\ee
Here, the normalized growth rate is defined as $D_+(z)\equiv D(z)/D(z_c)$, where $z_c$ is the collapse redshift (results are usually normalized 
to the observed low redshift quantities). Hence, choosing $z>z_c$ would allow us to test the validity of this approximation along the 
trajectory of the peak-patch at any time.
Furthermore,
\be
c_1^E(k,z_c) = \vk\cdot\vec L_\text{pk}^{(1)}(\vk) +  c_1(k,z_c) 
= \bigg\{ b_{10}+1+\Big[b_{01}-R_v^2\Big]k^2\bigg\}W_R(k)
\ee
is the first-order Eulerian peak bias.
Hence, our leading order result agrees with Eq.(\ref{eq:phhtree}), as expected.
Moreover, $P_\text{1-loop}(k,z_c)$ is the 1-loop mode-coupling term encapsulating the second-order contribution in the correlation functions,
\be
\label{eq:pmc_pkloc}
P_\text{1-loop}(k,z_c)= \frac12 (2\pi)^3  e^{-\frac{1}{3}k^2\sigma_{v,\text{pk}}^2}\int_{\vk_1} \int_{\vk_2}\!
\Bigl[c_2^E(\vk_1,\vk_2,z)\Big]^2 P_\text{L}(k_1)P_\text{L}(k_2)\,\ddir(\vk-\vk_1-\vk_2)\;.
\ee
The second-order Eulerian peak bias, 
\be
c_2^E(\vk_1,\vk_2,z) \equiv 2 {\cal F}_2(\vk_1,\vk_2) 
+ \Bigl[c_1(k_1,z_c){\cal F}_1(\vk_2)+c_1(k_2,z_c){\cal F}_1(\vk_1)\Bigr] 
+ c_2(\vk_1,\vk_2,z_c) \;,
\ee
\end{widetext}
is a linear combination of the first- and second-order Lagrangian peak biases $c_1(k,z_c)$ and $c_2(\vk_1,\vk_2,z_c)$, respectively, and 
involves the mode-coupling kernels
\begin{align}
{\cal F}_n(\vk_1,&\dots,\vk_n) \equiv 
\frac{1}{n!}\vk\cdot\vec L_\text{pk}^{(1)}(\vk_1) \times\, \cdots\, \times \vk\cdot\vec L_\text{pk}^{(1)}(\vk_n) \nonumber \\
&\equiv F_n(\vk_1,\dots,\vk_n)\, c_v(k_1)\times\,\cdots\,\times c_v(k_n) \;.
\label{eq:MCkernel}
\end{align}
Here, $F_n(\vk_1,\dots,\vk_n)$ are the usual (standard) perturbation theory mode-coupling kernels, which can be derived recursively 
\cite{goroff/grinstein/etal:1986,jain/bertschinger:1994,ptreview,rampf:2012}.
The presence of the filter $W_R(k)$ in Eq.(\ref{eq:MCkernel}) reflects the fact that the peak displacement field is assumed to be 
insensitive to the details of the mass distribution within a given Lagrangian peak - patch. The explicit expression of $c_2(\vk_1,\vk_2,z_c)$ 
and the corresponding second-order bias parameters can be found in, e.g., 
\cite{Desjacques:2012eb,moradinezhad/chan/etal:2016,matsubara/desjacques:2016} for a peak
constraint. Note that the assumption of spherical collapse implies that the second-order bias parameters $b_{K_2}$ associated with the 
Lagrangian tidal shear be zero. In the ellipsoidal collapse approximation however, $b_{K_2}$ is expected to be negative 
\cite{sheth/chan/scoccimarro:2013,castorina/paranjape/etal:2016}.

Of course, the mode-coupling is incorrect since we discarded higher-order displacements in LPT, so that the coefficients of the terms 
proportional to $(\kvh_1\cdot\kvh_2)^n\equiv \mu^n$ in the second-order kernel ${\cal F}_2$ are all equal to $1/2$. In the exact dynamics, 
these coefficients are $5/7$ (constant piece), $1/2$ ($\mu$-term) and $2/7$ ($\mu^2$-term) \cite{peebles:1980,fry:1984}.

Like semi-analytic methods which resum part of the perturbative expansion (such as e.g. RPT \cite{crocce/scoccimarro:2006a}), our Fourier 
space expansion Eq.(\ref{eq:xpkz1loop}) violates GI at any order (see \cite{peloso/pietroni:2013} for a detailed discussion) because
the exponential damping factor involves the dispersion of gradient modes $\partial_i\Phi$, where $\Phi$ is the gravitational potential.
As discussed above, constant, albeit possibly time-dependent gradient perturbations are unphysical because they can be removed through a
coordinate transformation \cite{scoccimarro/frieman:1996,kehagias/riotto:2013,peloso/pietroni:2013,Baldauf:2015xfa,dai/pajer/schmidt:2015}. 
However, since it is a perturbative expansion of the Zel'dovich peak correlation function, Eq.\ref{eq:xpkzexact}, it will necessarily 
satisfy GI on the scales where the expansion has converged towards the full Zel'dovich result. At 1-loop, the convergence is achieved at 
the percent level across the BAO. GI is obviously violated at short separations, but this is not a problem so long as one is interested
in modelling perturbatively halo statistics on mildly nonlinear scales. Solving Eq.(\ref{eq:xpkzexact}) numerically would provide a 
solution which satisfy GI at all scales. This is quite challenging in 3-dimensions but, as shown in \cite{baldauf/codis/etal:2016}, it is 
easily tractable in 1-dimension (where the Zel'dovich solution describes the exact dynamics \cite{buchert:1989}).

\subsubsection{Perturbative expansion in configuration space}
\label{sub:rspace}

The second approach begins with the evaluation of the Gaussian integral $\int_{\vk}$ over the wavemodes which, once it is performed, leaves 
us with a convolution over the Lagrangian separation of the form
\be
\label{eq:lagconv}
1 + \xpk^E(r,z) = \int\!d^3q\, F\big(\vq\lvert \vr\big) \;,
\ee
where the kernel
\begin{align}
F\big(&\vq\lvert\vr\big) = \frac{1}{\bnpk^2} \int\! d^N\! y_1 d^N\! y_2\,\npk(\bm{y}_1)\npk(\bm{y}_2)
p_2(\bm{y}_1,\bm{y}_2;\vq,z_i) \nonumber \\
&\times \frac{\exp\!\Big[-\frac12 (\vr-\vq-\Delta\vec\Psi)^\top\vcc^{-1}(\vr-\vq-\Delta\vec\Psi)\Big]}{\sqrt{(2\pi)^3{\rm det}\vcc}}
\label{eq:Fqr}
\end{align}
is related to the probability that a peak pair separated by a distance $\vr$ in the initial conditions has a separation $\vq$ at the time of 
halo collapse. Note that , unlike $\vcc$, $\Delta\vec\Psi$ has, among others, an explicit dependence on the peak height $\nu$ and the curvature 
$J_1\propto -{\rm tr}(\zeta_{ij})$.
Denoting the $n$th-order contributions to the joint covariance and mean displacement (subject to the peak constraint) as $\vcc^{(n)}$ and 
$\Delta\vec\Psi^{(n)}$, respectively, a first order expansion in correlation functions yields
\begin{widetext}
\begin{align}
\eh{-\frac12 (r-q-\Delta\Psi)_i\vcc^{-1}_{ij}(r-q-\Delta\Psi)_j} 
&\approx \eh{-\frac12 (r-q)_i\vcc^{(0),-1}_{ij}(r-q)_j} \\
&\qquad \times \left[1-\frac12 (r-q)_i\vcc^{(1),-1}_{ij}(r-q)_j+\Delta\Psi_i^{(1)}\vcc^{(0),-1}_{ij}(r-q)_j\right] \nonumber \;.
\end{align}
\end{widetext}
The advantage of having the mean displacements downstairs resides in the fact that we can take into account the peak constraint, i.e. average 
over the peak curvature $J_1$, to get $\Delta\overline{\vec\Psi}^{(1)}$. Once the latter is re-absorbed into the exponent, we obtain the 
following resumed expression,
\begin{align}
\label{eq:resumFqr}
F&\big(\vq\lvert\vr\big) = \Big[1+\xpk(q,z_i)\Big]  \\
&\quad \times \frac{\exp\!\Big[-\frac12 \big(\vr-\vq-\Delta\overline{\vec\Psi}^{(1)}\big)^\top\vcc^{-1}\big(\vr-\vq-\Delta\overline{\vec\Psi}^{(1)})\Big]}
{\sqrt{(2\pi)^3{\rm det}\vcc}}
\nonumber
\end{align}
which follows immediately from the fact that the first line in the right-hand side of Eq.~(\ref{eq:Fqr}) is precisely equal to $1+\xpk(q,z_i)$.
Since Eq.(\ref{eq:resumFqr}) is exact at first-order in correlation functions solely, deviations will arise at second- and higher-order  because 
$\overline{J_1^n}\ne \overline{J}_1^n$ does not hold for $n\geq 2$. As we will see below however, the 1-loop contribution is at most 5-10\% of
the tree-level contribution Eq.(\ref{eq:phhtree}) across the BAO (The typical shift around the BAO is $\sim 1\hmpc$ \cite{crocce/scoccimarro:2008}),
so that this approximation remains very good at large scales.

Note that, in the absence of any Lagrangian bias (so that $\Delta\vec\Psi^{(1)} \equiv 0$), we recover the well-known expression 
\be
F\big(\vq\lvert\vr\big)=\frac{1}{\sqrt{(2\pi)^3 \det\vcc}}\eh{-\frac12 (\vr-\vq)^\top\vcc^{-1}(\vr-\vq)}
\ee
for dark matter. Additional practical details on the evaluation of Eq.~\ref{eq:lagconv} can be found in Appendix \ref{app:za_rspace}.

Eq.(\ref{eq:lagconv}), together with the kernel Eq.(\ref{eq:resumFqr}), enables us to predict the evolved 2-point correlation for any separation $r$ 
once the Lagrangian correlation $\xpk(q,z_i)$ is known because the angular dependence on the azimuthal angle is trivial while that on the cosine 
$\mu=\qvh\cdot\rvh$ can be performed analytically (see Appendix \ref{app:za_rspace}).
Preliminary tests suggest that a numerical evaluation of $\xpk(q,z_i)$ is tractable for a moving, deterministic barrier, but more challenging when the 
barrier is fuzzy. 
For this reason, we will report results on this approach elsewhere, and focus here on the perturbative, Fourier space method described in 
Sec.~\ref{sub:kspace}.  

\subsection{Redshift Space Distortions}

Finally, let us emphasize that this phase space approach can be readily extended to redshift space, where the line-of-sight position is distorted by 
the peculiar velocity of the galaxy/halos relative to the Hubble flow. 
Following previous studies \cite{kaiser:1987,hivon/bouchet/etal:1995,fisher/nusser:1996,scoccimarro:2004,Matsubara:2011ck,uhlemann/kopp/haugg:2015}, 
the position in redshift space $\vec s$ in the distant-observer (or plane-parallel) approximation is given by 
\be
\vec s=\vec q+\vec \Psi+\big(\rvh\! \cdot\! \dot{\vec \Psi}\big)\ \rvh \equiv \vq + \mathcal{P}\, \vec\Psi \;,
\ee
where $\rvh$ is a unit vector in the line-of-sight direction. 
In the Zel'dovich approximation, the projection operator is simply  $\mathcal{P}=\text{I}+f\rvh\otimes\rvh$ (A similar projection operator 
$\mathcal{P}^{(n)}$ can be defined for all the higher-order displacements, as in \cite{Matsubara:2011ck}).
The time derivative in the above equation is straightforwardly performed yielding a logarithmic growth factor $f$.  This real-to-redshift space 
mapping implies that the covariance and mean of the relative displacement are modified by the following projections operators,
\be
\begin{split}
\Delta\Psi_i^s=&\Big(\delta_{il}^\text{(K)}+f \hat n_i \hat n_l\Big) \Delta\Psi_l\\
\vcc_{ij}^s=&\Big(\delta_{il}^\text{(K)}+f \hat n_i \hat n_l\Big)\vcc_{lm}\Big(\delta_{mj}^\text{(K)}+f \hat n_m \hat n_j\Big)
\end{split}
\ee
where a subscript $s$ denotes quantities in redshift space. Consequently, we shall for instance replace Eq.(\ref{eq:resumFqr}) by
\begin{align}
\label{eq:resumFqs}
F(\vec q|\vec s) &= \Big[1+\xpk(q;z_i)\Big]  \\
&\times \frac{\eh{-\frac12 (s-q- \Delta\bar \Psi^s)_i(\vcc^s)^{-1}_{ij}(s-q- \Delta\bar \Psi^s)_j}}
{\sqrt{(2\pi)^3 \det \vcc^s}} \nonumber \;.
\end{align}
These expressions extend the results of \cite{fisher/nusser:1996} to any Lagrangian bias. Of course, the Zel'dovich approximation misses late time 
small-scale motions induced by neighboring mass concentrations and, thus, provides a crude approximation to the redshift space power spectrum of
halo centers.

\subsection{Comparison with N-body simulations}

\begin{figure*}
\subfloat{\includegraphics[width=0.49\textwidth]{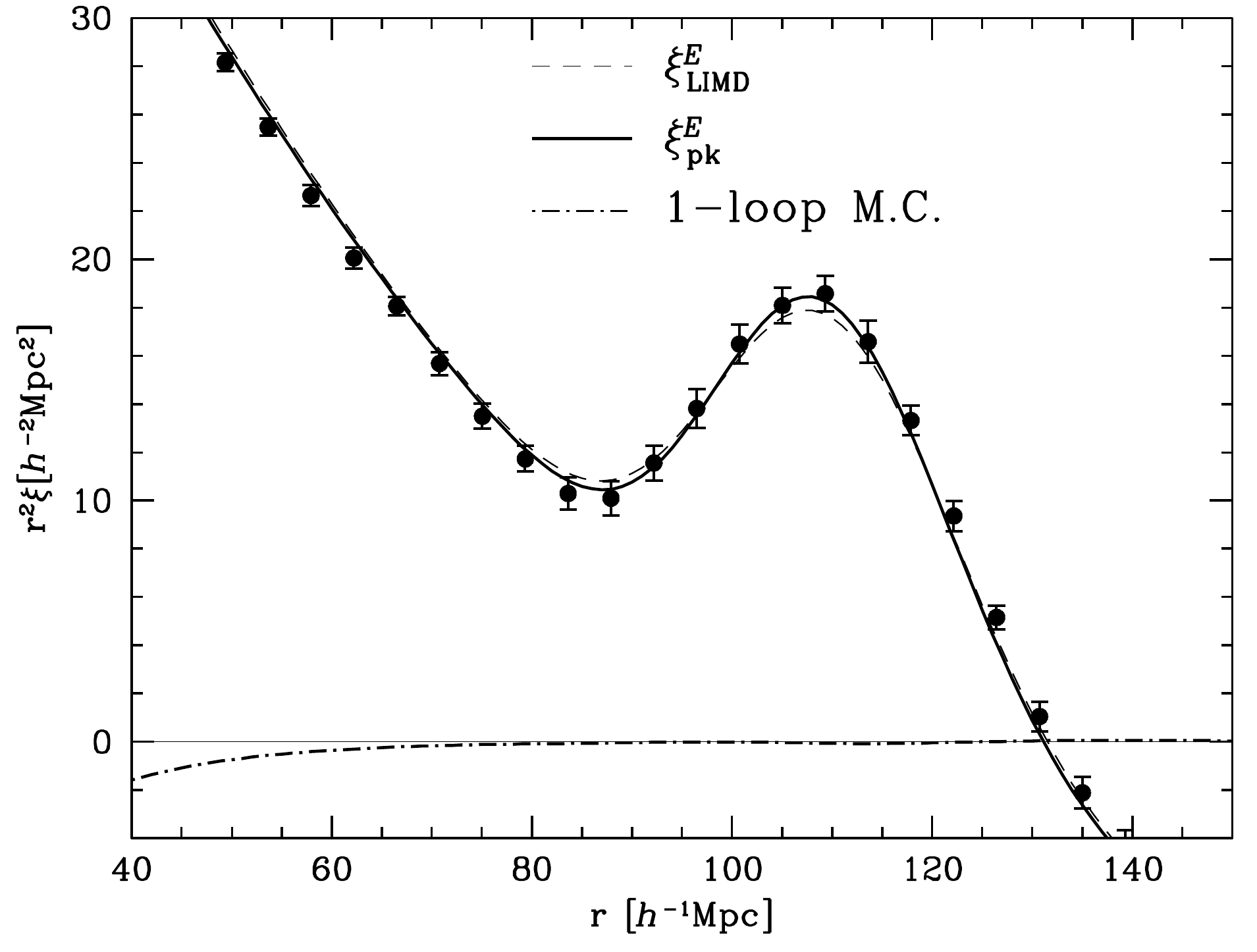}}
\subfloat{\includegraphics[width=0.49\textwidth]{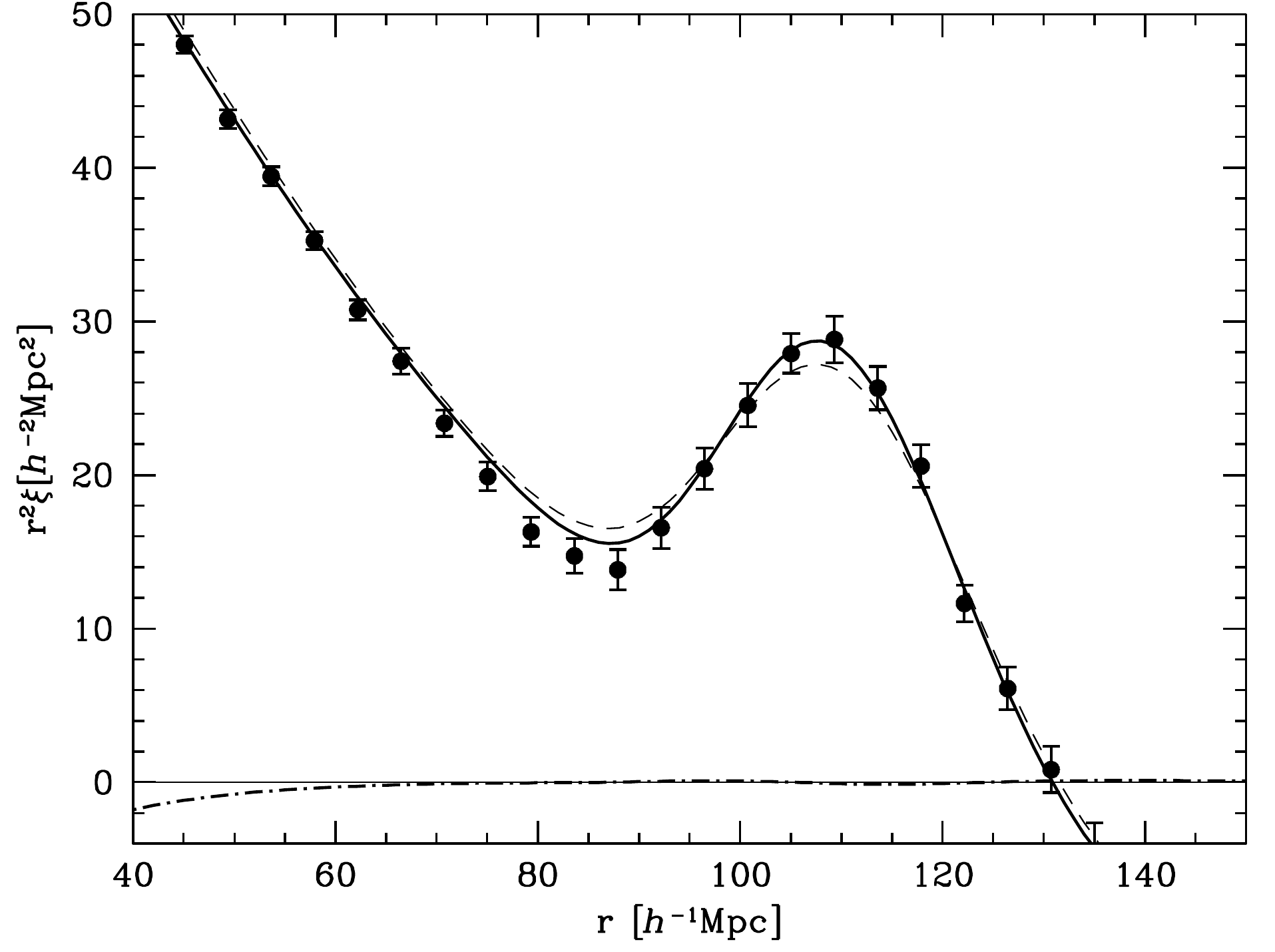}} \\
\subfloat{\includegraphics[width=0.49\textwidth]{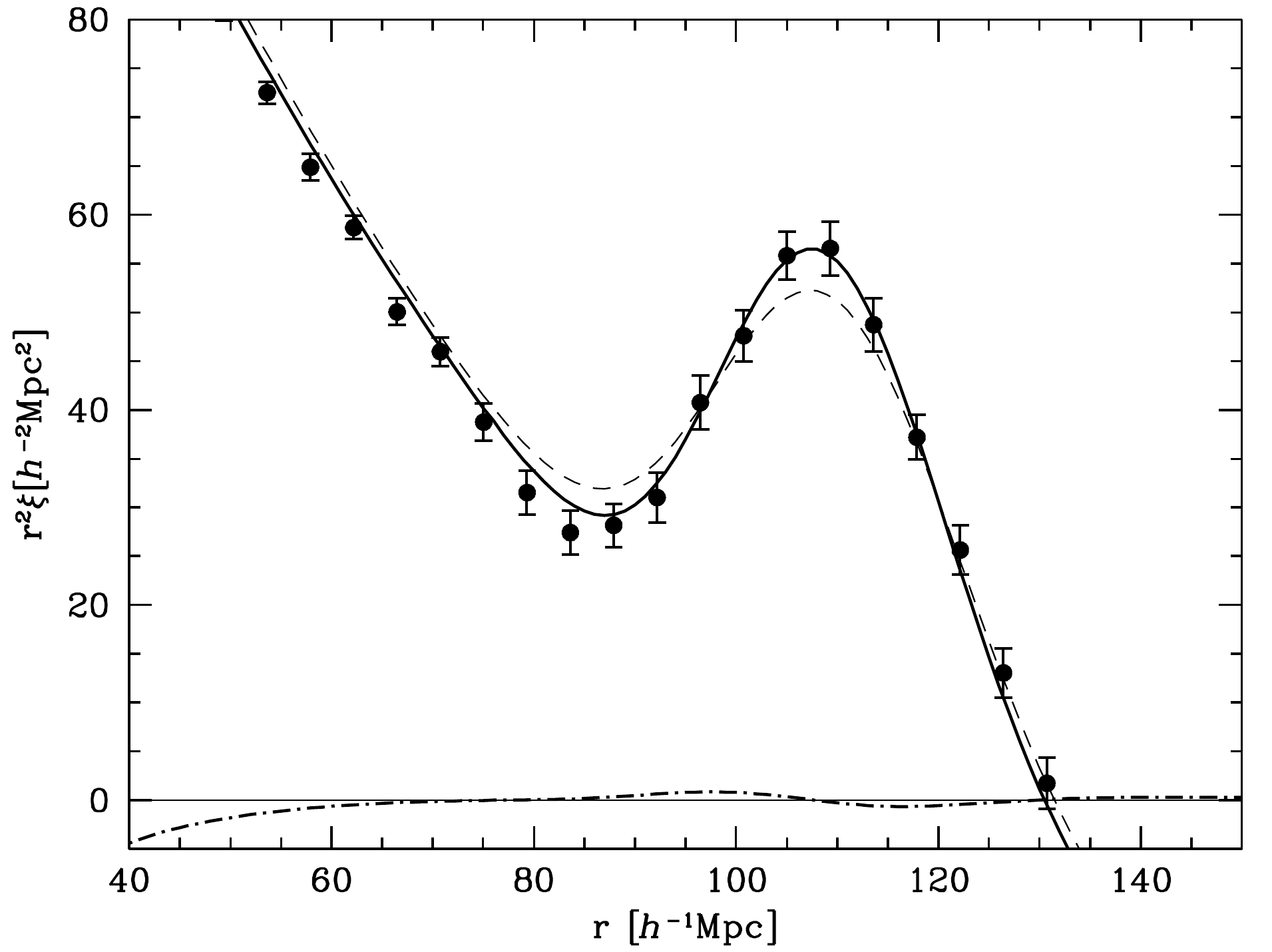}}
\subfloat{\includegraphics[width=0.49\textwidth]{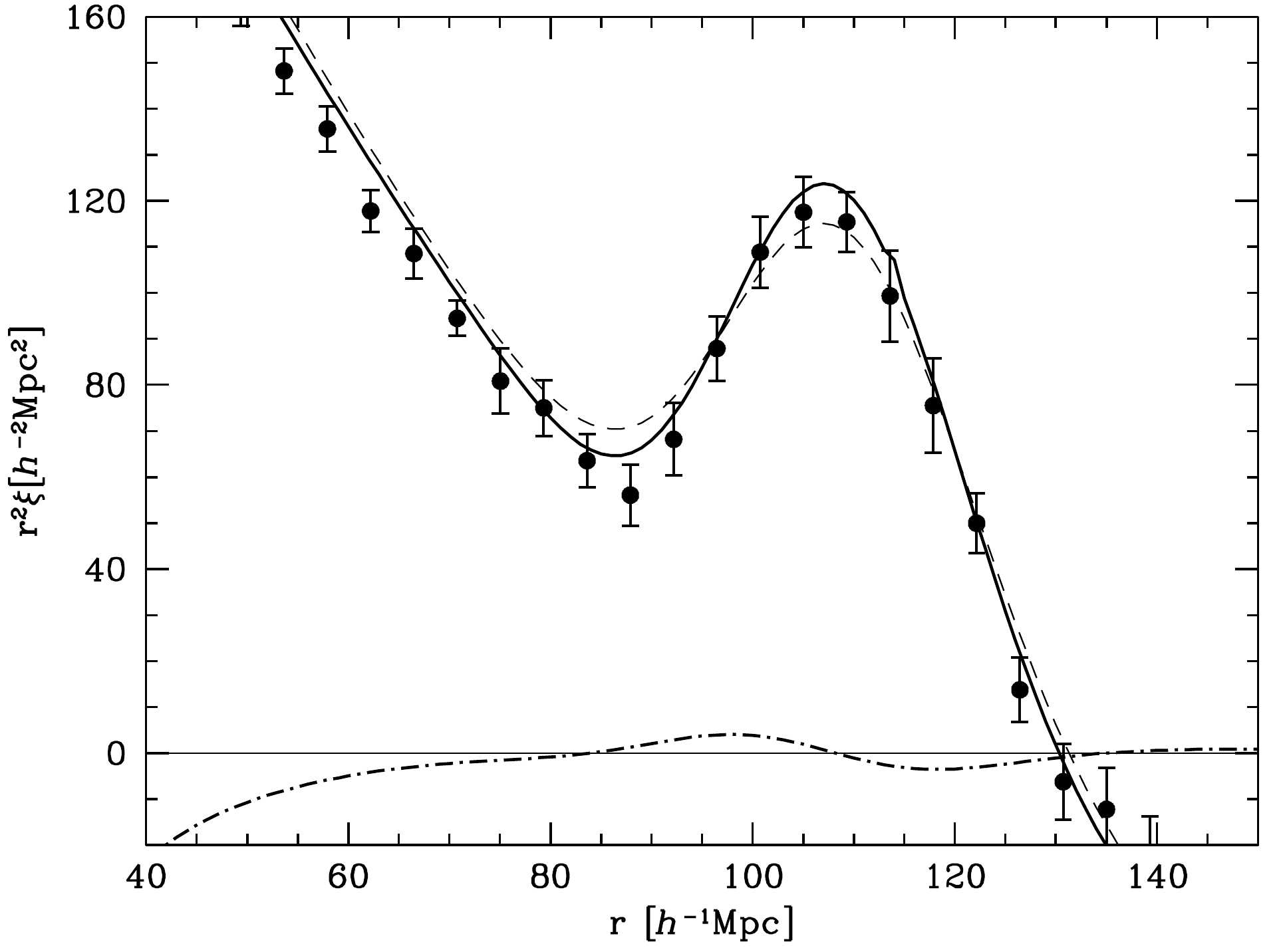}}
\caption{Enhancement of the BAO at redshift $z=0$ for the halo bins I -- IV from top left to bottom right. 
The continuous curve represents the peak model prediction, Eq.(\ref{eq:xpkz1loop}), in the Zel'dovich
approximation. The 1-loop mode-coupling contribution is shown as the dotted-dashed curve. Finally, the 
dashed curve is the prediction of a Lagrangian {\small LIMD} bias model. Eq.(\ref{eq:LIMD}) (see text). 
Note that none of the model parameters were fitted to the data shown in this figure. 
The peak model is a better match to the measurements, especially at high mass where the peak constraint is 
a good description of the Lagrangian halo patches.}
\label{fig:finalbao}
\end{figure*}

We are now in a position to compare theoretical predictions and measurements of the evolved halo correlation function.

In Fig.\ref{fig:finalbao}, the data points show the 2-point correlation function at redshift $z=0$ for the halo bins I -- IV. The solid curve is the
peak prediction Eq.(\ref{eq:xpkz1loop}) at 1-loop in the Zel'dovich approximation, whereas the dashed curve is a prediction based on a Lagrangian
{\small LIMD} bias scheme (an abbreviation for ``local in matter density'' in the terminology of \cite{biasreview}) or, more familiarly, the usual 
local Lagrangian bias model \cite{szalay:1988,fry/gaztanaga:1993,manera/gaztanaga:2011,schmidt/jeong/desjacques:2013}.
In the latter, the halo 2-point correlation function at the redshift of collapse $z_c$ takes the form
\begin{align}
\xi_\text{\tiny LIMD}^E(r,z_c) &= (1+b_{10})^2 \int_{\vk}\, e^{-\frac{1}{3}k^2\sigma_{v,\text{m}}^2} P_\text{L}(k) \nonumber \\
& \quad + P_\text{1-loop}(k,z) \;,
\label{eq:LIMD}
\end{align}
where the {\small LIMD} 1-loop mode-coupling power spectrum is given by
\begin{align}
P_\text{1-loop}(k,z) &= 2 e^{-\frac{1}{3}k^2\sigma_{v,\text{m}}^2} \int_{\vk_1}\int_{\vk_2}\,
\left[F_2(\vk_1,\vk_2) + b_{10}+\frac{b_{20}}{2}\right]^2 \nonumber \\
&\times P_\text{L}(k_1) P_\text{L}(k_2) \delta^\text{(D)}(\vk-\vk_1-\vk_2) \;.
\end{align}
The presence of $\sigma_{v,\text{m}}$ and $F_2(\vk_1,\vk_2)$ follows from the assumption of a vanishing velocity bias.
This model has already been considered in \cite{catelan/lucchin/etal:1998} (though they did not have the exponential damping factors), as well as 
\cite{crocce/scoccimarro:2008,smith/scoccimarro/sheth:2008} in the context of the BAO smearing and scale-dependence. 
Note that none of the parameters of both the peak and {\small LIMD} model were fitted to the $z=0$ data. Namely, once the collapse barrier has been
specified, the excursion set peak approach predicts a halo mass function $\bar n(M)$ (see e.g. \cite{Paranjape:2012jt}) from which we can derive all 
the bias parameters through a peak-background split \cite{Desjacques:2010gz,Desjacques:2012eb,Lazeyras:2015giz}. 
For this purpose, we assumed the collapse barrier to be of the square-root form, i.e. $B(\sigma_0)=\delta_c + \beta\sigma_0$ with $\beta$ lognormally 
distributed, where $\delta_c$ is the linear threshold for spherical collapse. 
This approximates well the moving barrier in the ellipsoidal collapse approximation \cite{moreno/giocoli/sheth:2009}. 
We then followed \cite{robertson/kravtsov/etal:2009} and calibrated the free parameters describing the barrier from the distribution of halo linear 
overdensities in the initial conditions. In particular, we found $\delta_c\approx 1.43$, a value significantly lower than the spherical collapse 
prediction $\delta_c\sim 1.68$. Having no free parameters left, we identified, for each halo bin, the mass $M$ corresponding to the bias $b_{10}$ 
inferred from the Lagrangian correlation function (Fig.\ref{fig:initialbao}) and computed all the remaining first- and second-order bias parameters 
from the peak-background split prescription of \cite{Desjacques:2012eb,Lazeyras:2015giz}. 
For the {\small LIMD} prediction Eq.(\ref{eq:LIMD}), we used the same values of $b_{10}$ and $b_{20}$ as in the peak model. 
In Table~\ref{table1}, we quote in parenthesis the values of $b_{01}$ and $R_v$ predicted by our excursion set peak model. The predictions are broadly
consistent with the measurements, albeit $\sim$ 30 -- 50\% larger for $b_{01}$ (resp. $\sim$ 10 -- 20\% larger for $R_v$) than the data. Note that the 
effective smoothing scale $k_*$ depends on the difference $R_v^2 - b_{01}<0$ (see Eq.~(\ref{eq:keffbaosmoothing})) and, hence, is only mildly affected by 
the systematic overestimation of $b_{01}$ and $R_v$. 

Fig.\ref{fig:finalbao} shows that the nonlinear gravitational evolution has significantly smeared out the BAO feature, so that the {\small LIMD} model 
compares much better with the $z=0$ data than in the initial conditions (see Fig.\ref{fig:initialbao}). Notwithstanding, the peak model furnishes a 
better fit to the $z=0$ data. Overall, the 1-loop mode-coupling contribution (only shown for the peak model as a dotted-dashed curve) is small, except
for the most massive bin where $b_{20}$ is noticeably larger than zero. Therefore, the predictions weakly depend on the exact form of $c_2(\vk_1,\vk_2,z_c)$.
Notice also that both the peak and LIMD prediction overestimate the data at separations $r\lesssim 80\hmpc$ for the most massive bins. 
We speculate that this may be due to the incorrect mode-coupling.
Our results are consistent with the findings of \cite{matsubara/desjacques:2016}, who found that the BAO enhancement in the final correlations varies
only at the few percent levels among different bias schemes. Note, however, that \cite{matsubara/desjacques:2016} did not follow a phase-space approach 
as in our study. Hence, our predictions differ from theirs even at tree-level (for instance, our exponential damping term depends on the peak velocity
dispersion rather than that of the matter).

\section{BAO reconstruction}
\label{sec:BAOrec}

Given the enhancement of the BAO signature in Lagrangian space relative to linear theory, and the good agreement between the peak predictions 
and the simulations at $z=z_i$ and $z=0$, one may wonder whether a naive reconstruction of the initial halo correlation function
(see, e.g., \cite{eisenstein/seo/etal:2007,kitaura/ensslin:2008,kitaura/hess:2013,jasche/wandelt:2013,schmittfull/feng/etal:2015,achitouv/blake:2015,
keselman/nusser:2016} for various implementations), in which 
the BAO feature is strongly enhanced by the curvature term $\propto b_{01}$, would increase the signal-to-noise ratio on the position of the BAO.
Unfortunately, reconstructing only the BAO in the halo sample decreases the overall amplitude of the signal since the initial and evolved halo 
correlation functions, $\xi_\text{hh}$ and $\xhh$, are related at linear level through (ignoring any possible contribution from higher derivative 
bias)
\be
\xi_\text{hh}(r)=\left(\frac{b_{10}}{D(z_i)}\right)^2D(z_i)^2\xi_\text{L}(r)=\left(b_{10}\right)^2\xi_\text{L}(r) \;.
\ee
Here, $\xi_\text{L}(r)$ is the linear matter correlation function (suitably smoothed on the Lagrangian halo scale).
Therefore, 
\begin{equation}
\xi_\text{hh}=
\left(\frac{b_{10}}{1+b_{10}}\right)^2\xhh\equiv\left(\frac{b_{10}^\text{(E)}-1}{b_{10}^\text{(E)}}\right)^2\xhh \;.
\end{equation}
This enhances the relative importance of sampling variance and thus the relative error. 
One possible remedy would be to perform a partial reconstruction, i.e. up to a certain redshift $z < z_i$ that optimizes both the amplitude of the
broadband and the contrast of the BAO.

Alternatively, one may also consider a random sample of particles and displace both the halos and the random particles 
(see, e.g., \cite{padmanabhan/xu/etal:2012,kazin/koda/etal:2014,burden/percival/etal:2014} for recent applications). The discussion of
\cite{noh/white/padmanabhan:2009,padmanabhan/white/cohn:2009} straightforwardly generalizes to discrete objects, such that the density field 
of the displaced halo (peak) and random tracers read
\begin{align}
\delta_\text{d}(\vk) &= \frac{1}{\bnpk} \int\! d^3q\,e^{-i\vk\cdot(\vq+\vec\Psi+\vs)} \sum_\text{pk}\delta^{(\text{D})}\!(\vq-\vq_\text{pk}) \\
\delta_\text{s}(\vk) &= \frac{1}{\bar n_\text{rnd}} \int\! d^3q\,e^{-i\vk\cdot(\vq+\vs)} \sum_\text{rnd}\delta^{(\text{D})}\!(\vq-\vq_\text{rnd}) \;,
\end{align}
where the subscripts ``d'' and ``s'' stand for ``displaced'' and ``shifted'', as in \cite{padmanabhan/white/cohn:2009}. 
Deplacing the random sample is crucial because this produces an anti-biased version of the matter density field. Therefore, although the
overall amplitude of the displaced 2-point function $\xi_\text{dd}(r)$ decays from $(1+b_{10})^2$ to $b_{10}^2$, the amplitude of the 
``reconstructed'' field $\delta_\text{r}(\vk)\equiv \delta_\text{d}(\vk)-\delta_\text{s}(\vk)$ is still $(1+b_{10})^2$. 
More precisely, let the Fourier modes of the inverse, reconstructed displacement field be
\be
\label{eq:recondisp}
\vec s(\vk) \equiv -i \frac{\kvh}{k} \mathcal{S}(k)\,\delta_\text{L}(\vk) \;,
\ee
where $\mathcal{S}(k)$ is a low-pass filter with a characteristic smoothing scale $R_\mathcal{S}$. The filter $\mathcal{S}(k)$ could, in principle, 
be optimized for the BAO reconstruction (see \cite{vargas/ho/etal:2015} for a recent numerical investigation of the effect of smoothing on the 
reconstruction). 
The reconstructed power spectrum $P_\text{rr}(\vk)$ is given by
\be
P_\text{rr}(\vk) = P_\text{dd}(\vk)-2 P_\text{ds}(\vk)+P_\text{ss}(\vk) \;,
\ee
where, at tree-level, the auto- and cross-power spectra are given by
\begin{align}
P_\text{dd}(\vk) &= e^{-\frac{1}{3}k^2\sigma_\text{dd}^2} \Big[c_v(k)-\mathcal{S}(k)+c_1(k,z_c)\Big]^2 P_\text{L}(k) \nonumber \\
P_\text{ds}(\vk) &= -e^{-\frac{1}{3}k^2\sigma_\text{ds}^2}\, \mathcal{S}(k)  \Big[c_v(k)-\mathcal{S}(k)+c_1(k,z_c)\Big] P_\text{L}(k) \nonumber \\
P_\text{ss}(\vk) &= e^{-\frac{1}{3}k^2\sigma_\text{ss}^2}\,\mathcal{S}^2(k) P_\text{L}(k) \;.
\end{align}
where the dispersions of the displaced and shifted field are given by
\begin{align}
\sigma_\text{dd}^2 &= \int_{\vk} k^{-2}\big(c_v(k)-\mathcal{S}(k)\big)^2 P_\text{L}(k) \\
\sigma_\text{ss}^2 &= \int_{\vk} k^{-2} \mathcal{S}^2(k) \,P_\text{L}(k) \;,
\end{align}
whereas $\sigma_\text{ds}^2 = (\sigma_\text{dd}^2+\sigma_\text{ss}^2)/2$.
To understand how the reconstruction affects the BAO damping when the bias features a $k^2$ dependence at linear level, we proceed in analogy 
with the effective BAO smoothing scale introduced in Sec.~\ref{sec:keff} and collect all the $k^2$ correction terms to the reconstructed power 
spectrum. Upon factorizing the constant piece of the linear, {\it Eulerian} bias, we can write
\be
\label{eq:reconprr}
P_\text{rr}(\vk)\approx \big(1+b_{10}\big)^2\Bigg[1-\bigg(\frac{k}{k_*}\bigg)^2\Bigg] P_\text{L}(k) \;, 
\ee 
where the effective BAO smoothing scale $k_*$ of the reconstructed field is given by
\be
k_*= \Bigg(\frac{1}{3}\frac{\sigma_\text{ss}^2+ b_{10}\sigma_\text{dd}^2}{1+b_{10}}  
+R^2+2\frac{R_v^2-b_{01}}{1+b_{10}}\Bigg)^{-1/2} \;.
\ee
Eq.(\ref{eq:reconprr}) makes explicit that the broadband power spectrum of the reconstructed field has the same amplitude as that of the 
data.
Sharpening the BAO peak corresponds to increasing the value of $k_*$. In general, the optimal choice of $\mathcal{S}$ will depend on the 
exact value of $b_{10}$, $b_{01}$ and $R$. In the limit where the tracers are unbiased however, $b_{10}=0$ and $k_*$ is maximized when  
$\sigma_\text{ss}\ll 1$, i.e. when the smoothing scale $R_\mathcal{S}$ is large. Note that the term $R_v^2-b_{01}$ still contributes to the 
broadening in this regime. Conversely, in the limit of highly biased tracers, we have $b_{10}\gg 1$ which, upon using the relation 
$b_{01}=(\sigma_0/\sigma_1)^2 (\nu^2/\delta_c-b_{10})$ yields
\be
k_* = \left(\frac{1}{3}\sigma_\text{dd}^2 + R^2\right)^{-1/2} \;.
\ee
The effective smoothing scale is now maximized with the choice $\mathcal{S}(k) \sim c_v(k) = (1- R^2 k^2) W_R(k)$, i.e.  a reconstruction which 
takes into account the velocity bias in the halo displacement field. Of course, one should consider higher-order terms in the development, in 
which case a large value of $R_\mathcal{S}$ may not be the optimal limit in the unbiased case.

\begin{figure}
\includegraphics[width=0.5\textwidth]{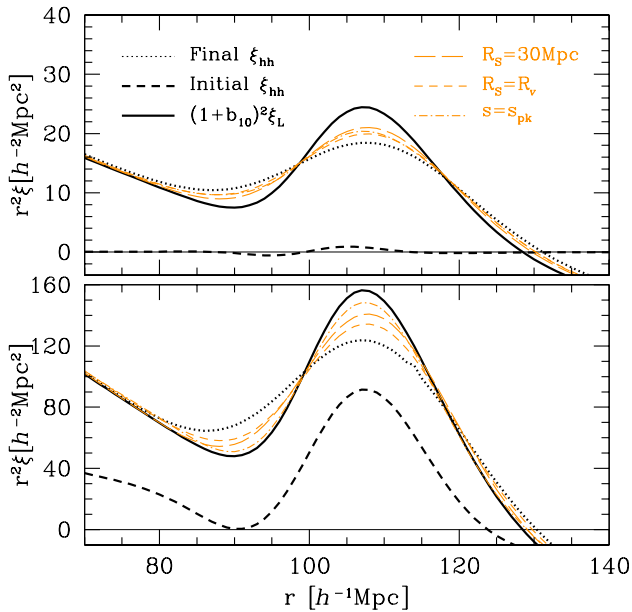}
\caption{The $z=0$ and initial halo correlation function (dotted and dashed black curve) for the mass bin I (top panel) and IV (bottom panel).
The thin (orange) curves show the correlation after a Zel'dovich reconstruction using a Gaussian kernel with filtering scale  $R_S=30\hMpc$ 
(long-dashed) and $R_S=R_v$ (short-dashed), and a reconstructed displacement field $s_\text{pk}$ similar to the biased peak displacement
(see text). While there is no significant difference between the various reconstruction schemes at low mass, using  $s_\text{pk}$ appears to 
perform better for the high mass bin.}
\label{fig:recon}
\end{figure}

To illustrate our discussion, we show in Fig.\ref{fig:recon} examples of reconstruction of the $z=0$ peak correlation function $\xi_\text{pk,f}$ 
(dotted curve) corresponding to the least and most massive halo bin (top and bottom panel, respectively). The solid curve represent the unsmoothed, 
linear theory correlation multiplied by a factor of $(1+b_{10})^2$. Since we expect $k_*>0$ (note the factor of $R_v^2-b_{01}$ though), this 
corresponds to the best possible reconstruction in the context of a displaced and a shifted field. The solid dashed curve indicates the initial
peak correlation $\xi_\text{pk,i}$, which is strongly suppressed for low mass halos with $b_{10}\sim 0$.

We consider 3 different reconstructed displacement field, Eq.(\ref{eq:recondisp}): a Gaussian filtering kernel $S(k) = \exp(-k^2 R_S^2/2)$ with 
radius i) $R_S=30\hMpc$ (thin long-dashed) and ii) $R_S=R_v$ (thin short-dashed) together with a filtering kernel $S(k)=(1-R_v^2 k^2)\exp(-k^2R_v^2/2)$
which capture part of the linear displacement or velocity peak bias $c_v(k)$. At low mass, the reconstruction work best for a large value of $R_S$,
as advertised above. Including the velocity bias $c_v$ does not make much of a difference because $R_v$ is much smaller than the width $\sim 10-15
\hMpc$ of the BAO. At high mass however where the Lagrangian halo scale $R$ is a significant fraction of the BAO width, the filtering kernel
$S(k)=(1-R_v^2 k^2)\exp(-k^2R_v^2/2)$ yields the best reconstruction. We have not explored this issue any further here, but it would be interesting
to assess whether this may have practical applications for BAO reconstruction scheme. We leave this to future work.

\section{Conclusions}
\label{sec:conclusion}

We have extended the analysis of \cite{Desjacques:2010gz} in several directions. Firstly, unlike \cite{Desjacques:2010gz} who assumed the BBKS peak
constraint to describe Lagrangian halos, our predictions rely on the more realistic excursion set peak approach. This has been shown to furnish a
good fit to the mass function and LIMD bias parameters of halos \cite{Paranjape:2012jt,paranjape/sefusatti/etal:2013,lazeyras/wagner/etal:2016}.
Secondly, we have demonstrated using N-body simulations that the BAO in the initial, 2-point correlation function of halos is strongly enhanced by 
the curvature term $\bm{\nabla}^2\delta_\text{L}$, in agreement with the prediction of \cite{Desjacques:2008jj} and with the Fourier space analysis 
of \cite{elia/ludlow/porciani:2012,Baldauf:2015ve}. This is true for all the halos considered here, i.e. $M\geq 7.8\times 10^{12}\hmsun$.
The simple Lagrangian LIMD bias model is found to be inconsistent with these measurements. 
Thirdly, we have shown that a Lagrangian perturbation theory approach -- restricted to the Zel'dovich approximation for simplicity -- can consistently 
evolve the initial, Lagrangian data to the final, Eulerian space if one allows for a linear halo velocity bias. Peak theory predicts a linear halo
velocity dispersion in very good agreement with the data, although it underestimates the dispersion of the total halo displacement presumably because 
2LPT and higher-order displacements are missing. In the evolved, 2-point halo correlation, the smoothing of the BAO is very similar to that of the 
dark matter, so that the LIMD approximation without both the higher-derivative bias $\bm{\nabla}^2\delta_\text{L}$ and the velocity bias does not fall 
very far from the data points. The halo BAO smoothing is nonetheless a bit weaker than that of the matter owing to the fact that 
i) $\sigma_{v,\text{pk}}< \sigma_{v,\text{m}}$ and ii) $R_v^2-b_{01} < 0$. 
Finally, we have also pointed out that a reconstruction of the BAO may be improved if the Zel'dovich peak displacement -- which include the velocity
bias -- is used instead of the matter displacement corresponding to random field points. 
We intend to investigate this issue in the future, along with the effect of a $\bm{\nabla}^2\delta_\text{L}$ -- selection on the BAO contrast.
Namely, since the contrast of the BAO is sensitive to the value of $b_{01} \sim b_{\bm{\nabla}^2\delta}^L$, and since the latter (or, equivalently, the
density of the ``large-scale'' environment) correlates with the formation history of the halo 
(see, e.g., \cite{zentner:2007,dalal/white/etal:2008,desjacques:2008}), we expect the BAO of halos of a given mass $M$ to be enhanced or suppressed if 
they are selected according to their formation time for instance.
We will investigate this issue in future work.

\section*{Acknowledgement}

We would like to thank Matias Zaldarriaga for useful discussions. 
V.D. acknowledges partial support from the Swiss National Science Foundation, the Boninchi Foundation and 
the Israel Science Foundation (grant no. 1395/16).

\bibliographystyle{prsty}
\bibliography{bao}

\appendix

\begin{widetext}

\section{Peaks in the Zel'dovich approximation}
\label{app:za}

The Eulerian position of peaks is related to the Lagrangian peaks by 
\be
\begin{split}
1+\delta_\text{pk}(\vec r)
&=\frac{1}{\bnpk}\sum_\text{pk} \ddir\left(\vr- \vec r_\text{pk}\right)
=\frac{1}{\bnpk}\int\! d^3q'\, \ddir\left[ \vr- \vec q'- \sigma_{v,\text{m}}\vec \Psi(\vec q')\right]\sum_\text{pk}\ddir( \vec q'- \vec q_\text{pk})\,,\\
=&\frac{1}{\bnpk}\int\! d^3q'\, \int_{\vk}\eh{\ii  \vk \scalp ( \vr-\vec  q')}\sum_\text{pk}\ddir( \vec q'- \vec q_\text{pk})
\eh{-\ii \sigma_{v,\text{m}} \vk \scalp \vec \Psi( \vec q')} \;,
\end{split}
\ee
where the normalized, linear displacement $\vec\Psi$ is evaluated at the redshift $z_c$ of halo collapse. 
We can now calculate the correlation function
\be
\label{eq:xpkz0}
1+\xpk^E(r)=\la\delta_\text{pk}(\vec 0)\delta_\text{pk}(\vec r)\ra
=\frac{1}{\bnpk^2}\int\! d^3q \int_{\vk} \eh{\ii \vec k\scalp (\vr - \vq)}
\Bigl\langle \exp\Big[-\ii \sigma_{v,\text{m}}\vec k\scalp(\vec \Psi_2-\vec \Psi_1)\Big]  \npk(\bm{y}_1) \npk(\bm{y}_2)\Bigr\rangle\; .
\ee
Here, $\vq=\vec q_2-\vec q_1$ is the Lagrangian separation of the peaks, $\vec \Psi_1$ and $\vec \Psi_2$ are the displacements at their respective 
Lagrangian positions and $\npk(\bm{y}_i)$ are the peak selection functions. 
For BBKS peaks \cite{Bardeen:1985tr}, the peak state vector $\bm{y}=(\nu,\eta_i,\zeta_A)$ involves the peak height $\nu$, the components $\eta_i$ of 
the first derivative of the smoothed density $\delta_R$, and the 6 independent components $\zeta_A$ of the Hessian 
$\zeta_{ij}\propto \partial_i\partial_j\delta_R$. All these variables are normalized, such that $\la\nu^2\ra=1$, $\la\eta_i\eta_j\ra=(1/3)\delta_{ij}^\text{(K)}$
etc. Here, $\delta_{ij}^\text{(K)}$ is the Kronecker symbol.
The ensemble average in the right-hand side of Eq.(\ref{eq:xpkz0}) is performed over the peak state vectors $\bm{y}_i$ and the peak displacements 
$\vec\Psi_i$. Marginalizing over the latter, we eventually arrive at Eq.(\ref{eq:xpkzexact}).

\section{Second order derivation: Fourier space}
\label{app:za_kspace}

We shall only outline the key steps of the derivation of the 1-loop contribution since it is fairly similar to that given in \cite{Desjacques:2010gz}.
Decomposing the displacement field onto the helicity basis $(\ve_+,\qvh,\ve_-)$ (this is the usual Helmholtz decomposition, see e.g. \cite{chan:2014}), 
such that $\qvh\cdot \ve_{\pm}=0$, the covariance matrix $\vcc$ becomes diagonal, i.e. $\vcc={\rm diag}(C_{+},C_{0},C_{-})$, where $C_{m}$ are complex 
numbers, and the mean vector reads $\Delta\vec\Psi=(\Delta\Psi_{+},\Delta\Psi_{0},\Delta\Psi_{-})$. 
We can now solve for $C_m$ and $\Delta\Psi_m$ perturbatively in terms of the linear power spectrum $P_\text{L}(k)$. Namely
\begin{align}
C_m(\vq) &= \sum_{n\geq 0} C_m^{(n)}(\vq) 
= C_m^{(0)}+\sum_{n\geq 1} \int_{\vk_1}\dots\int_{\vk_n}\, C_m^{(n)}(\vk_1,\dots,\vk_n;\vq)\, e^{i(\vk_1+\dots+\vk_n)\cdot\vq} \\
\Delta\Psi_m(\vq) &= \sum_{n\geq 0}\Delta\Psi_m^{(n)}(\vq)
= \Delta\Psi_m^{(0)}+\sum_{n\geq 1} \int_{\vk_1}\dots\int_{\vk_n}\, \Delta\Psi_m^{(n)}(\vk_1,\dots,\vk_n;\vq)\, e^{i(\vk_1+\dots+\vk_n)\cdot\vq} \;.
\end{align}
Here, $C_m^{(n)}(\vk_1,\dots,\vk_n;\vq)$ and $\Delta\Psi_m^{(n)}(\vk_1,\dots,\vk_n;\vq)$ involves exactly $n$ factors of $P_\text{L}(k_i)$. 
The extra dependence on $\vq$ ensures that both $C_m^{(n)}$ and $\Delta\Psi_m^{(n)}$ can depend on the angles between the Fourier wavevectors $\vk_i$ and the 
separation vector $\vq$.
We emphasize that, without any constraint (i.e. for random field points), all the terms with $n\geq 2$ vanish. Therefore, their presence reflects the 
complicated dependence of $\vcc$ and $\Delta\vec\Psi$ on the separation $q$ induced by the constraint.
At zeroth order, the components of the covariance matrix are given by
\be
C_{m}^{(0)}=\frac{1}{3}\left(1-\gamma_v^2\right)
\ee
where $\gamma_v\equiv \sigma_0^2/(\sigma_{-1}\sigma_1)$ quantifies the strength of the correlation between the (normalized) peak displacement and the 
gradient $\partial_i\delta_R$. At this order however, there is no net flow, such that $\Delta\Psi_m^{(0)}=0$. Therefore,
\be
e^{-\frac{1}{2}\sigma_{v,\text{m}}^2\vk^\top\vcc\vk+i\sigma_{_{v,\text{m}}}\vk^\top\cdot\Delta\vec\Psi}\approx
e^{-\frac{1}{3}\sigma_{v,\text{pk}}^2 k^2} \;.
\ee
At the first order, the covariance matrices $\vcc_m$ receive a contribution whose Fourier transform reads
\begin{align}
C_0^{(1)}(\vk_1;\vq) &=-\frac{2}{\sigma_{-1}^2} \frac{(\qvh\cdot\kvh_1)^2}{k_1^2}\, c_v^2(k_1) P_\text{L}(k_1) \\
C_{\pm}^{(1)}(\vk_1;\vq) &= -\frac{2}{\sigma_{-1}^2} \frac{(\ve_{\pm}\cdot\kvh_1)(\ve_{\pm}^*\cdot\kvh_1)}{k_1^2}\, c_v^2(k_1) P_\text{L}(k_1)
\nonumber \;.
\end{align}
where the linear peak velocity bias $c_v(k)$ is defined in Eq.(\ref{eq:bv}), and $\ve_{\pm}^*$ is the complex conjugate of $\ve_{\pm}$.
There is also a non-vanishing flow along the line-of-sight, 
\begin{equation}
\label{eq:Xi0_1st}
\Delta\tilde{\Psi}_0^{(1)}(\vk_1;\vq) = \frac{2i}{\sigma_{-1}}
\frac{(\qvh\cdot\kvh_1)}{k_1} c_v(k_1) \left(b_\nu+b_{J_1} k_1^2\right)W_R(k_1) P_\text{L}(k_1) \;.
\end{equation}
The tilde indicates that we have already averaged out the anisotropies in the peak density profile (they are proportional to the traceless components 
of $\zeta_{ij}$). Moreover, the linear peak bias parameters $b_\nu$ and $b_{J_1}$ of peaks of height $\nu$ and curvature $J_1$ are given by
\be
b_\nu = \frac{1}{\sigma_0} \left(\frac{\nu-\gamma_1 J_1}{1-\gamma_1^2}\right) \;,\qquad
b_{J_1} = \frac{1}{\sigma_2}\left(\frac{J_1-\gamma_1\nu}{1-\gamma_1^2}\right) \;.
\ee
They yield $b_{10}$ and $b_{01}$ once averaged over the peak constraint.
By contrast, the transverse (or divergence-free) components $\Delta\tilde{\Psi}_{\pm}^{(1)}$ vanish after the averaging. Expanding the terms involving 
$P_\text{L}(k)$ up to first order, but keeping up the zero-lag contribution $\propto \sigma_{v,\text{pk}}^2 k^2$, we arrive at 
\be
\label{eq:interO1}
1+\xpk^E(r,z_c) = \int_{\vk} e^{-\frac{1}{3}\sigma_{v,\text{pk}}^2 k^2}\int\! d^3q\ e^{\ii \vk\cdot(\vr-\vq)}
\biggl\langle \bigg(1-\frac{1}{2}\sigma_{v,\text{m}}^2\vk^\top\vcc^{(1)}\vk-i\sigma_{v,\text{m}}\vk^\top\Delta\vec\Psi^{(1)}\bigg) 
\big(1+\xpk(q)\big)\bigg\rangle + \mathcal{O}(P_\text{L}^2)\;,
\ee
where $\xpk(q)$ is the Lagrangian peak correlation function. To illustrate how this expression can be simplified, let us consider the term
\begin{equation}
\int_{\vk} e^{-\frac{1}{3}\sigma_{v,\text{pk}}^2 k^2}\int\! d^3q\ e^{\ii \vk\cdot(\vr-\vq)}
\biggl\langle \bigg(-i\sigma_{v,\text{m}}\vk^\top\Delta\vec\Psi^{(1)}(\vq)\bigg)\bigg\rangle \;.
\end{equation}
Since $\Delta\tilde\Psi_{\pm}^{(1)}=0$, this simplifies to
\begin{align}
-i\sigma_{v,\text{m}}&\int_{\vk} e^{-\frac{1}{3}\sigma_{v,\text{pk}}^2 k^2}\int\! d^3q\ e^{\ii \vk\cdot(\vr-\vq)}
\int_{\vk_1}\Big\langle k_0\Delta\Psi_0^{(1)}(\vk_1;\vq)\Big\rangle e^{\ii\vk_1\cdot\vq} \\
&= 2\frac{\sigma_{v,\text{m}}}{\sigma_{-1}}\int_{\vk} e^{-\frac{1}{3}\sigma_{v,\text{pk}}^2 k^2}e^{\ii\vk\cdot\vr}
\int_{\vk_1}\,\left(\frac{k}{k_1}\right)c_v(k_1)\left(b_{10}+b_{01}k_1^2\right)W_R(k_1)P_\text{L}(k_1)
\int\! d^3q\, (\qvh\cdot\kvh)(\qvh\cdot\kvh_1)\,e^{\ii(\vk_1-\vk)\cdot\vq} \nonumber \\
&= 2 \int_{\vk} e^{-\frac{1}{3}\sigma_{v,\text{pk}}^2 k^2} c_v(k)  \Big(b_{10}+b_{01}k^2\Big)W_R(k)P_\text{L}(k) e^{\ii\vk\cdot\vr} 
\nonumber \;,
\end{align}
where we have imposed the peak constraint, and successively used $k_0\equiv k(\qvh\cdot\kvh)$, $\sigma_{v,\text{m}}\equiv \sigma_{-1}$, 
and the relation
\begin{equation}
\int\!d^3q\,\qh_i\qh_j\, e^{\ii(\vk-\vk_1)\cdot\vq} = (2\pi)^3 \delta^{\text{(D)}}\!(\vk-\vk_1) \,\kh_i \kh_j \;.
\end{equation}
After similar manipulations, Eq.(\ref{eq:interO1}) eventually leads to Eq.(\ref{eq:xpkz1loop}).
Note the difference with the iPT, which resums only the contributions proportional to $\sigma_{v,\text{m}}^2k^2$ into the exponential, 
such that the exponential damping piece is proportional to the dark matter velocity dispersion \cite{Matsubara:2011ck}.

The second-order or 1-loop contribution to the peak correlation function is more involved but, again, straightforward to evaluate once all the terms 
involving correlations at finite separations have been expanded. Summarizing, the 1-loop mode coupling term is given by i) the second-order contribution to 
the Lagrangian correlation function $\xpk(q)$; ii) the terms proportional to $(\vcc^{(1)})^2$, $\vcc^{(1)}\Delta\vec\Psi^{(1)}$, $(\Delta\vec\Psi^{(1)})^2$ plus the
first-order contribution to $\xpk(q)$ times $\Delta\vec\Psi^{(1)}$ and $\vcc^{(1)}$;  and iii) terms involving $\vcc^{(2)}$ and $\Delta\vec\Psi^{(2)}$. 
In particular, we have
\begin{align}
C_0^{(2)}(\vk_1,\vk_2;\vq) &= \frac{1}{\sigma_{-1}^2}\Biggl\{-\frac{3}{\sigma_1^2}
(\qvh\cdot\kvh_1)(\qvh\cdot\kvh_2)+\frac{5}{4\sigma_2^2}k_1 k_2
\left[3(\qvh\cdot\kvh_1)^2-1\right]\left[3(\qvh\cdot\kvh_2)^2-1\right] \\
&\qquad +\left(1-\gamma_1^2\right)^{-1}\left[\frac{1}{\sigma_0^2}
(k_1 k_2)^{-1}+\frac{1}{\sigma_2^2}k_1 k_2-\frac{2\gamma_1}
{\sigma_0\sigma_2}k_1^{-1} k_2\right]\Biggr\}(\qvh\cdot\kvh_1)
(\qvh\cdot\kvh_2) \nonumber \\ 
&\qquad\times c_v(k_1) c_v(k_2) W_R(k_1) W_R(k_2) P_\text{L}(k_1)P_\text{L}(k_2) \nonumber \\
C_{\pm}^{(2)}(\vk_1,\vk_2;\vq) &= \frac{1}{\sigma_{-1}^2}
\left[-\frac{3}{\sigma_1^2}+\frac{15}{\sigma_2^2}k_1 k_2(\qvh\cdot\kvh_1)
(\qvh\cdot\kvh_2)\right](\ve_{\pm}\cdot\kvh_1)(\ve_{\pm}^*\cdot\kvh_1)
(\ve_{\pm}\cdot\kvh_2)(\ve_{\pm}^*\cdot\kvh_2) \\
&\qquad\times c_v(k_1) c_v(k_2) W_R(k_1) W_R(k_2) P_\text{L}(k_1) P_\text{L}(k_2) \nonumber
\end{align}
whereas, for the line-of-sight components of the mean displacement, we have
\begin{align}
\Delta\tilde{\Psi}_0^{(2)}(\vk_1,\vk_2;\vq) &= \frac{2i}{\sigma_{-1}}
\Biggl\{\left(1-\gamma_1^2\right)^{-1}\Biggl(\frac{1}{\sigma_0^2}q_1^{-1}
+\frac{1}{\sigma_2^2}k_1 k_2^2 -\frac{\gamma_1}{\sigma_0\sigma_2} 
k_1^{-1} k_2^2-\frac{\gamma_1}{\sigma_0\sigma_2}k_1\Biggr)
+\frac{3}{\sigma_1^2}k_2(\qvh\cdot\kvh_1)(\qvh\cdot\kvh_2) \\ 
& \quad -\frac{5}{4\sigma_2^2}k_1 k_2^2\left[3(\qvh\cdot\kvh_1)^2-1\right]
\left[3(\qvh\cdot\kvh_2)^2-1\right]\Biggr\}(\qvh\cdot\kvh_1)
c_v(k_1)\left(b_\nu+b_{J_1} k_2^2\right) W_R(k_1)W_R^2(k_2)P_\text{L}(k_1) P_\text{L}(k_2) \nonumber \;.
\end{align}
Once again, the contribution $\Delta\tilde\Psi_{\pm}^{(2)}$ vanishes when the peak constraint is taken into account. Of course, this should 
come up as no surprise since the displacement field is curl-free in the Zel'dovich approximation \cite{Zeldovich:1970,shandarin/zeldovich:1989}. 
The peak constraint does not affect this property, so that we expect $\Delta\Psi_{\pm}^{(n)}\equiv 0$ at any order $n$.

Since the calculation is a bit lengthy, let us focus on the pieces proportional to $1/\sigma_1^2$ in $C_m^{(2)}$ for illustration.
Their contribution to the 1-loop mode-coupling is through the second-order term $-(1/2)\vk^\top\vcc^{(2)}\vk$. 
Performing the integral
\be
\int\! d^3q\ e^{-\ii \vk\cdot\vq}\bigg\langle \bigg(-\frac{1}{2}\sigma_{v,\text{m}}^2\vk^\top\vcc^{(2)}\vk\bigg)\bigg\rangle
= -\frac{1}{2}\sigma_{v,\text{m}}^2\int\! d^3q\,e^{-\ii \vk\cdot\vq}\left(k_0\big\langle C_0^{(2)}\big\rangle k_0
+ \sum_{a=\pm} k_a\big\langle C_a^{(2)} \big\rangle k_a^*\right) 
\ee
and extracting the contribution proportional to $1/\sigma_1^2$ gives, on using $k_{\pm}\equiv \vk\cdot\ve_{\pm}$ and $k_{\pm}^*\equiv\vk\cdot\ve_{\pm}^*$,
\begin{align}
\frac{3}{2\sigma_1^2}\int\!d^3q\,e^{-\ii\vk\cdot\vq} \int_{\vk_1} \int_{\vk_2}
& \biggl[(\qvh\cdot\vk)^2(\qvh\cdot\kvh_1)^2(\qvh\cdot\kvh_2)^2 + \sum_{a=\pm}
(\ve_a\cdot\vk)(\ve_a^*\cdot\vk)(\ve_a\cdot\kvh_1)(\ve_a^*\cdot\kvh_1)
(\ve_a\cdot\kvh_2)(\ve_a^*\cdot\kvh_2)\biggr]  \nonumber \\ 
& \times c_v(k_1) c_v(k_2) W_R(k_1) W_R(k_2) P_\text{L}(k_1) P_\text{L}(k_2) e^{\ii(\vk_1+\vk_2)\cdot\vq}\;.
\label{eq:chi1cont1}
\end{align}
To make progress, let us momentarily ignore the factors of $c_v(k)$ and focus on the square brackets. We have
\begin{multline}
\int_{\vk_1} \int_{\vk_2}
\biggl[(\qvh\cdot\vk)^2(\qvh\cdot\kvh_1)^2(\qvh\cdot\kvh_2)^2 + \sum_{a=\pm}
(\ve_a\cdot\vk)(\ve_a^*\cdot\vk)(\ve_a\cdot\kvh_1)(\ve_a^*\cdot\kvh_1)
(\ve_a\cdot\kvh_2)(\ve_a^*\cdot\kvh_2)\biggr] W_R(k_1) W_R(k_2) P_\text{L}(k_1) P_\text{L}(k_2) e^{\ii(\vk_1+\vk_2)\cdot\vq} \\
= \frac{1}{9}
\biggl[(\qvh\cdot\vk)^2\left(\xi_0^{(0)}(q)-2\xi_2^{(0)}(q)\right)^2
+\left(\xi_0^{(0)}(q)+\xi_2^{(0)}(q)\right)^2\sum_{a=\pm}(\ve_a\cdot\vk)(\ve_a^*\cdot\vk)\biggr] \;.
\end{multline}
The functions
\be
\xi_\ell^{(n)}(r) \equiv \frac{1}{2\pi^2}\int_0^\infty\!dk\,k^{2(n+1)} W_R^2(k) P_\text{L}(k)j_\ell(kr) \;,
\ee
where $j_\ell(x)$ are spherical Bessel functions, generalizes the spectral moments to finite separations. 
To proceed further, we use 
\be
\label{eq:int_rr}
\frac{1}{4\pi}\int\!\!d\Omega_{\kvh}\,\kh_i\kh_j\,e^{i\vk\cdot\vq}
=\frac{1}{3}\Bigl[j_0(kq)+j_2(kq)\Bigr]\delta_{ij}-j_2(kq)\,\qh_i\qh_j\;,
\ee
and observe that
\begin{align}
\frac{1}{(4\pi)^2}\int\!\!d\Omega_{\kvh_1}\int\!\!d\Omega_{\kvh_2}&\,
(\vk\cdot\kvh_1)(\vk\cdot\kvh_2)
(\kvh_1\cdot\kvh_2)\,e^{i(\vk_1+\vk_2)\cdot\vq} \\ 
& =\delta_{j\alpha}k_i k_\beta \frac{1}{(4\pi)^2}
\int\!\!d\Omega_{\kvh_1}\int\!\!d\Omega_{\kvh_2}\,
\kh_{1i}\kh_{1j}\kh_{2\alpha}\kh_{2\beta} \,e^{i(\vk_1+\vk_2)\cdot\vq} \nonumber \\
&= \delta_{j\alpha}k_1 k_\beta
\left[\frac{1}{3}\Bigl(j_0(k_1q)-2j_2(k_1q)\Bigr)\qh_i\qh_j
+\frac{1}{3}\Bigl(j_0(k_1q)+j_2(k_1q)\Bigr)
\sum_{a=\pm}e_{ai}e_{aj}^*\right] \nonumber \\
&\qquad \times
\left[\frac{1}{3}\Bigl(j_0(k_2q)-2j_2(k_2q)\Bigr)\qh_\alpha\qh_\beta
+\frac{1}{3}\Bigl(j_0(k_2q)+j_2(k_2q)\Bigr)
\sum_{a=\pm}e_{a\alpha}e_{a\beta}^*\right] \nonumber \\
&= \frac{1}{9}
\biggl[(\qvh\cdot\vk)^2\Bigl(j_0(k_1q)-2j_2(k_1q)\Bigr)\Bigl(j_0(k_2q)-2j_2(k_2q)\Bigr)
+\Bigl(j_0(k_1q)+j_2(k_1q)\Bigr) \nonumber \\
& \qquad \times \Bigl(j_0(k_2q)+j_2(k_2q)\Bigr)\sum_{a=\pm}(\ve_a\cdot\vk)(\ve_a^*\cdot\vk)\biggr]
\nonumber \;.
\end{align}
To obtain the last equality, we took advantage of the fact that $\qvh\cdot\ve_{\pm}=\qvh\cdot\ve_\pm^*=0$.
Therefore, we have the following relation
\begin{equation}
\biggl[(\qvh\cdot\vk)^2(\qvh\cdot\kvh_1)^2(\qvh\cdot\kvh_2)^2 + \sum_{a=\pm}
(\ve_a\cdot\vk)(\ve_a^*\cdot\vk)(\ve_a\cdot\kvh_1)(\ve_a^*\cdot\kvh_1)
(\ve_a\cdot\kvh_2)(\ve_a^*\cdot\kvh_2)\biggr]=(\vk\cdot\kvh_1)(\vk\cdot\kvh_2)
(\kvh_1\cdot\kvh_2) \;,
\end{equation}
so that Eq.(\ref{eq:chi1cont1}) turns out to be equal to
\begin{align}
\frac{3}{2\sigma_1^2}\int\!d^3q\,e^{-\ii\vk\cdot\vq}
\int_{\vk_1}\int_{\vk_2} (\vk\cdot\kvh_1)(\vk\cdot\kvh_2)&(\kvh_1\cdot\kvh_2)\,
c_v(k_1) c_v(k_2) W_R(k_1) W_R(k_2) P_\text{L}(k_1) P_\text{L}(k_2) e^{i(\vk_1+\vk_2)\cdot\vq} \\
&\equiv \int_{\vk_1}\int_{\vk_2}\bigg[ {\cal F}_2(\vk_1,\vk_2) \Big(\chi_1\, W_R(k_1) W_R(k_2)\Big) \bigg] 
P_\text{L}(k_1)P_\text{L}(k_2) \delta^{\text{(D)}}\!(\vk-\vk_1-\vk_2) \nonumber \;.
\end{align}
Here, $\chi_1=-3/(2\sigma_1^2)$ is the Lagrangian peak bias factor associated to the normalized, chi-square variable 
$\eta^2(\vx)\equiv (\partial_i\delta_R)^2/\sigma_1^2$ \cite{Desjacques:2012eb,Lazeyras:2015giz}. 
This bias has been found to be different from zero for halos with mass $M\gtrsim M_\star$ \cite{biagetti/etal:2014}.
Hence, this term is a piece of the product ${\cal F}_2(\vk_1,\vk_2) c_2(\vk_1,\vk_2,z_c)$ that appear in Eq.(\ref{eq:xpkz1loop}).
The simplification of the other terms proceeds analogously.

\section{Second order derivation: Configuration space}
\label{app:za_rspace}

To perform the integral over $d^3q$ in Eq.(\ref{eq:resumFqr}), we decompose $\vr$ onto the helicity basis $(\ve_+,\qvh,\ve_-)$ 
and write $\qvh\cdot\vr=r\mu$, $\sum_{a=\pm}(\ve_a\cdot\vr)(\ve_a^*\cdot\vr)=r^2(1-\mu^2)$, so that the argument of the 
exponential can be written
\begin{align}
\big(\vr-\vq-\Delta\ov{\vec\Psi}^{(1)}\big)^\top\vcc^{-1}\big(\vr-\vq-\Delta\ov{\vec\Psi}^{(1)}\big) &=
\big(r\mu-q-\Delta\ov{\Psi}_0^{(1)}\big)^2 (\vcc^{-1})_\parallel + r^2 (1-\mu^2) (\vcc^{-1})_\perp \\
&= \big(q+\Delta\ov{\Psi}_0^{(1)}\big)^2(\vcc^{-1})_\parallel
+r^2(\vcc^{-1})_\perp - 2r\big(q+\Delta\ov{\Psi}_0^{(1)}\big)^2(\vcc^{-1})_\parallel \mu \nonumber \\
&\qquad +r^2\Big[(\vcc^{-1})_\parallel-(\vcc^{-1})_\perp\Big]\mu^2 \nonumber \;,
\end{align}
since there are no net motions transverse to $\qvh$ at linear order, $\Delta\ov{\vec\Psi}_{\pm}^{(1)}\equiv 0$. 
Here, $(\vcc^{-1})_\parallel \equiv (\vcc^{-1})_0$ and $(\vcc^{-1})_\perp\equiv (\vcc^{-1})_{\pm}$. 
As in \cite{taylor/hamilton:1996} (who considered unbiased tracers), the integral over the azimuthal angle $\phi$ trivially yields 
$2\pi$, while the integral over the cosine $\mu=\qvh\cdot\rvh$ of the polar angle can be performed analytically using 
\begin{equation}
\int_{-1}^1 \derd \mu \eh{-\frac12 \alpha \mu^2+\beta \mu+\gamma}=
{\rm exp}\!\left(\frac{\beta^2}{2 \alpha} + 
  \gamma\right) \sqrt{\frac{\pi}{2\alpha}} \left(\text{erf}\left[\frac{\alpha - \beta}{\sqrt{2 \alpha}}\right] + 
   \text{erf}\left[\frac{\alpha + \beta}{\sqrt{2 \alpha}}\right]\right) \;,
\end{equation}
leaving only one scalar numerical integral over the magnitude of the Lagrangian separation $q$, which both $\xpk(q,z_i)$ and
det$\vcc$ depend on. We could also keep the rms displacement dispersion $\vcc$ un-expanded at the non-perturbative level. This 
is motivated by the fact that $\xi_0^{(-1)}/\sigma_{-1}^2$ is only suppressed by a factor of $\mathcal{O}(10^1)$ at the BAO scale, 
while the other Lagrangian correlators $\xi_0^{(n)}$ with $n\geq 0$ are down by a factor of $\mathcal{O}(10^2)$.

\end{widetext}

\end{document}

%% file: defs.tex
\renewcommand{\vec}{\bm}
\newcommand{\be}{\begin{equation}}
\newcommand{\ee}{\end{equation}}
\newcommand{\derd}{\text{d}}
\newcommand{\eh}[1]{\exp\left[#1\right]}
\newcommand{\ii}{\text{i}}
\newcommand{\la}{\left\langle}
\newcommand{\ra}{\right\rangle}
\newcommand{\scalp}{\!\cdot\!}
\newcommand{\hMpc}{h^{-1}\text{Mpc}}

\newcommand{\ddir}{\delta^\text{(D)}}

\def\lsim{~\rlap{$<$}{\lower 1.0ex\hbox{$\sim$}}}
\def\bsim{~\rlap{$>$}{\lower 1.0ex\hbox{$\sim$}}}

\def\hmpc{\ {\rm {\it h}^{-1}Mpc}}

\def\hmsun{\ {\rm M_\odot/{\it h}}}

\newcommand{\ov}{\overline}
\newcommand{\xpk}{\xi_\text{pk}}
\newcommand{\npk}{n_\text{pk}}
\newcommand{\bnpk}{\bar{n}_\text{pk}}
\newcommand{\ve}{\vec e}
\newcommand{\vk}{\vec k}
\newcommand{\vq}{\vec q}
\newcommand{\vs}{\vec s}
\def\vr{\vec r}
\newcommand{\vx}{\vec x}

\def\kh{\mathit{\hat{k}}}
\def\qh{\mathit{\hat{q}}}

\newcommand{\kvh}{\hat{\bm{k}}}
\newcommand{\qvh}{\hat{\bm{q}}}
\newcommand{\rvh}{\hat{\bm{r}}}
\newcommand{\vcc}{\mathrm{C}}
\newcommand{\xhh}{\xi_\text{hh}^E}
\newcommand{\bnh}{\bar n_\text{h}}

\def\apjl{Astrophys.\ J.\ Lett.}

\def\aap{Astron.\ Astrophys.}

\def\apjs{Astrophys.\ J. Supp.}

%% file: bao.bbl
\begin{thebibliography}{10}

\bibitem{peebles/yu:1970}
P.~J.~E. {Peebles} and J.~T. {Yu}, \apj {\bf 162},  815  (1970).

\bibitem{sunyaev/zeldovich:1970}
R.~A. {Sunyaev} and Y.~B. {Zeldovich}, \apss {\bf 7},  3  (1970).

\bibitem{cole/2df:2005}
S. {Cole}, W.~J. {Percival}, J.~A. {Peacock}, P. {Norberg}, C.~M. {Baugh},
  C.~S. {Frenk}, I. {Baldry}, J. {Bland-Hawthorn}, T. {Bridges}, R. {Cannon},
  M. {Colless}, C. {Collins}, W. {Couch}, N.~J.~G. {Cross}, G. {Dalton}, V.~R.
  {Eke}, R. {De Propris}, S.~P. {Driver}, G. {Efstathiou}, R.~S. {Ellis}, K.
  {Glazebrook}, C. {Jackson}, A. {Jenkins}, O. {Lahav}, I. {Lewis}, S.
  {Lumsden}, S. {Maddox}, D. {Madgwick}, B.~A. {Peterson}, W. {Sutherland}, and
  K. {Taylor}, \mnras {\bf 362},  505  (2005).

\bibitem{eisenstein/sdss:2005}
D.~J. {Eisenstein}, I. {Zehavi}, D.~W. {Hogg}, R. {Scoccimarro}, M.~R.
  {Blanton}, R.~C. {Nichol}, R. {Scranton}, H.-J. {Seo}, M. {Tegmark}, Z.
  {Zheng}, S.~F. {Anderson}, J. {Annis}, N. {Bahcall}, J. {Brinkmann}, S.
  {Burles}, F.~J. {Castander}, A. {Connolly}, I. {Csabai}, M. {Doi}, M.
  {Fukugita}, J.~A. {Frieman}, K. {Glazebrook}, J.~E. {Gunn}, J.~S. {Hendry},
  G. {Hennessy}, Z. {Ivezi{\'c}}, S. {Kent}, G.~R. {Knapp}, H. {Lin}, Y.-S.
  {Loh}, R.~H. {Lupton}, B. {Margon}, T.~A. {McKay}, A. {Meiksin}, J.~A.
  {Munn}, A. {Pope}, M.~W. {Richmond}, D. {Schlegel}, D.~P. {Schneider}, K.
  {Shimasaku}, C. {Stoughton}, M.~A. {Strauss}, M. {SubbaRao}, A.~S. {Szalay},
  I. {Szapudi}, D.~L. {Tucker}, B. {Yanny}, and D.~G. {York}, \apj {\bf 633},
  560  (2005).

\bibitem{Ludlow:2011pk}
A.~D. {Ludlow} and C. {Porciani}, \mnras {\bf 413},  1961  (2011).

\bibitem{bond/myers:1996a}
J.~R. {Bond} and S.~T. {Myers}, \apjs {\bf 103},  1  (1996).

\bibitem{bond/myers:1996b}
J.~R. {Bond} and S.~T. {Myers}, \apjs {\bf 103},  41  (1996).

\bibitem{bond/myers:1996c}
J.~R. {Bond} and S.~T. {Myers}, \apjs {\bf 103},  63  (1996).

\bibitem{Bardeen:1985tr}
J.~M. Bardeen, J.~R. Bond, N. Kaiser, and A.~S. Szalay, Astrophys. J. {\bf
  304},  15  (1986).

\bibitem{fry/gaztanaga:1993}
J.~N. {Fry} and E. {Gaztanaga}, \apj {\bf 413},  447  (1993).

\bibitem{Desjacques:2008jj}
V. Desjacques, Phys. Rev. {\bf D78},  103503  (2008).

\bibitem{Desjacques:2010gz}
V. Desjacques, M. Crocce, R. Scoccimarro, and R.~K. Sheth, Phys. Rev. {\bf
  D82},  103529  (2010).

\bibitem{epstein:1983}
R.~I. {Epstein}, \mnras {\bf 205},  207  (1983).

\bibitem{bond/cole/etal:1991}
J.~R. {Bond}, S. {Cole}, G. {Efstathiou}, and N. {Kaiser}, \apj {\bf 379},  440
   (1991).

\bibitem{appel/jones:1990}
L. {Appel} and B.~J.~T. {Jones}, \mnras {\bf 245},  522  (1990).

\bibitem{elia/ludlow/porciani:2012}
A. {Elia}, A.~D. {Ludlow}, and C. {Porciani}, \mnras {\bf 421},  3472  (2012).

\bibitem{Baldauf:2015ve}
T. Baldauf, V. Desjacques, and U. Seljak, Phys. Rev. {\bf D92},  123507
  (2015).

\bibitem{paranjape/sheth:2012}
A. {Paranjape} and R.~K. {Sheth}, \mnras {\bf 426},  2789  (2012).

\bibitem{Paranjape:2012jt}
A. Paranjape, R.~K. Sheth, and V. Desjacques, Mon. Not. Roy. Astron. Soc. {\bf
  431},  1503  (2013).

\bibitem{noh/white/padmanabhan:2009}
Y. {Noh}, M. {White}, and N. {Padmanabhan}, \prd {\bf 80},  123501  (2009).

\bibitem{biasreview}
V. {Desjacques}, D. {Jeong}, and F. {Schmidt}, ArXiv e-prints  (2016).

\bibitem{Desjacques:2009kt}
V. Desjacques and R.~K. Sheth, Phys. Rev. {\bf D81},  023526  (2010).

\bibitem{biagetti/desjacques/etal:2016}
M. {Biagetti}, V. {Desjacques}, A. {Kehagias}, D. {Racco}, and A. {Riotto},
  \jcap {\bf 4},  040  (2016).

\bibitem{modi/castorina/seljak:2016}
C. {Modi}, E. {Castorina}, and U. {Seljak}, ArXiv e-prints  (2016).

\bibitem{Carlson:2013co}
J. {Carlson}, B. {Reid}, and M. {White}, \mnras {\bf 429},  1674  (2013).

\bibitem{Matsubara:2011ck}
T. Matsubara, Phys. Rev. {\bf D83},  083518  (2011).

\bibitem{Zeldovich:1970}
Y.~B. {Zel'dovich}, \aap {\bf 5},  84  (1970).

\bibitem{Matsubara:2007wj}
T. Matsubara, Phys. Rev. {\bf D77},  063530  (2008).

\bibitem{buchert:1989}
T. {Buchert}, \aap {\bf 223},  9  (1989).

\bibitem{hivon/bouchet/etal:1995}
E. {Hivon}, F.~R. {Bouchet}, S. {Colombi}, and R. {Juszkiewicz}, \aap {\bf
  298},  643  (1995).

\bibitem{moutarde/alimi/etal:1991}
F. {Moutarde}, J.-M. {Alimi}, F.~R. {Bouchet}, R. {Pellat}, and A. {Ramani},
  \apj {\bf 382},  377  (1991).

\bibitem{Bouchet:1994xp}
F.~R. Bouchet, S. Colombi, E. Hivon, and R. Juszkiewicz, Astron. Astrophys.
  {\bf 296},  575  (1995).

\bibitem{Scoccimarro:1997gr}
R. Scoccimarro, Mon. Not. Roy. Astron. Soc. {\bf 299},  1097  (1998).

\bibitem{Gunn:1972sv}
J.~E. Gunn and J.~R. Gott, III, Astrophys. J. {\bf 176},  1  (1972).

\bibitem{davis/peebles:1977}
M. {Davis} and P.~J.~E. {Peebles}, \apjs {\bf 34},  425  (1977).

\bibitem{bharadwaj:1996}
S. {Bharadwaj}, \apj {\bf 472},  1  (1996).

\bibitem{Baldauf:2015xfa}
T. Baldauf, M. Mirbabayi, M. Simonovic, and M. Zaldarriaga, Phys. Rev. {\bf
  D92},  043514  (2015).

\bibitem{Prada:2014bra}
F. Prada, C.~G. Scoccola, C.-H. Chuang, G. Yepes, A.~A. Klypin, F.-S. Kitaura,
  and S. Gottlober,   (2014).

\bibitem{rosen:1972}
G. Rosen, Am. J. Phys. {\bf 40},  683  (1972).

\bibitem{scoccimarro/frieman:1996}
R. {Scoccimarro} and J. {Frieman}, \apjs {\bf 105},  37  (1996).

\bibitem{kehagias/riotto:2013}
A. {Kehagias} and A. {Riotto}, Nuclear Physics B {\bf 873},  514  (2013).

\bibitem{peloso/pietroni:2013}
M. {Peloso} and M. {Pietroni}, \jcap {\bf 5},  031  (2013).

\bibitem{kehagias/norena/etal:2014}
A. {Kehagias}, J. {Nore{\~n}a}, H. {Perrier}, and A. {Riotto}, Nuclear Physics
  B {\bf 883},  83  (2014).

\bibitem{valageas:2014}
P. {Valageas}, \prd {\bf 89},  083534  (2014).

\bibitem{goroff/grinstein/etal:1986}
M.~H. {Goroff}, B. {Grinstein}, S.-J. {Rey}, and M.~B. {Wise}, \apj {\bf 311},
  6  (1986).

\bibitem{jain/bertschinger:1994}
B. {Jain} and E. {Bertschinger}, \apj {\bf 431},  495  (1994).

\bibitem{ptreview}
F. {Bernardeau}, S. {Colombi}, E. {Gazta{\~n}aga}, and R. {Scoccimarro},
  \physrep {\bf 367},  1  (2002).

\bibitem{rampf:2012}
C. {Rampf}, \jcap {\bf 12},  004  (2012).

\bibitem{Desjacques:2012eb}
V. Desjacques, Phys. Rev. {\bf D87},  043505  (2013).

\bibitem{moradinezhad/chan/etal:2016}
A. {Moradinezhad Dizgah}, K.~C. {Chan}, J. {Nore{\~n}a}, M. {Biagetti}, and V.
  {Desjacques}, \jcap {\bf 9},  030  (2016).

\bibitem{matsubara/desjacques:2016}
T. {Matsubara} and V. {Desjacques}, \prd {\bf 93},  123522  (2016).

\bibitem{sheth/chan/scoccimarro:2013}
R.~K. {Sheth}, K.~C. {Chan}, and R. {Scoccimarro}, \prd {\bf 87},  083002
  (2013).

\bibitem{castorina/paranjape/etal:2016}
E. {Castorina}, A. {Paranjape}, O. {Hahn}, and R.~K. {Sheth}, ArXiv e-prints
  (2016).

\bibitem{peebles:1980}
P.~J.~E. {Peebles}, {\em Research supported by the National Science
  Foundation.~Princeton, N.J., Princeton University Press, 1980.~435 p.}
  (PUBLISHER, ADDRESS, 1980).

\bibitem{fry:1984}
J.~N. {Fry}, \apj {\bf 279},  499  (1984).

\bibitem{crocce/scoccimarro:2006a}
M. {Crocce} and R. {Scoccimarro}, \prd {\bf 73},  063519  (2006).

\bibitem{dai/pajer/schmidt:2015}
L. {Dai}, E. {Pajer}, and F. {Schmidt}, \jcap {\bf 11},  043  (2015).

\bibitem{baldauf/codis/etal:2016}
T. {Baldauf}, S. {Codis}, V. {Desjacques}, and C. {Pichon}, \mnras {\bf 456},
  3985  (2016).

\bibitem{crocce/scoccimarro:2008}
M. {Crocce} and R. {Scoccimarro}, \prd {\bf 77},  023533  (2008).

\bibitem{kaiser:1987}
N. {Kaiser}, \mnras {\bf 227},  1  (1987).

\bibitem{fisher/nusser:1996}
K.~B. {Fisher} and A. {Nusser}, \mnras {\bf 279},  L1  (1996).

\bibitem{scoccimarro:2004}
R. {Scoccimarro}, \prd {\bf 70},  083007  (2004).

\bibitem{uhlemann/kopp/haugg:2015}
C. {Uhlemann}, M. {Kopp}, and T. {Haugg}, \prd {\bf 92},  063004  (2015).

\bibitem{szalay:1988}
A.~S. {Szalay}, \apj {\bf 333},  21  (1988).

\bibitem{schmidt/jeong/desjacques:2013}
F. {Schmidt}, D. {Jeong}, and V. {Desjacques}, \prd {\bf 88},  023515  (2013).

\bibitem{manera/gaztanaga:2011}
M. {Manera} and E. {Gazta{\~n}aga}, \mnras {\bf 415},  383  (2011).

\bibitem{catelan/lucchin/etal:1998}
P. {Catelan}, F. {Lucchin}, S. {Matarrese}, and C. {Porciani}, \mnras {\bf
  297},  692  (1998).

\bibitem{smith/scoccimarro/sheth:2008}
R.~E. {Smith}, R. {Scoccimarro}, and R.~K. {Sheth}, \prd {\bf 77},  043525
  (2008).

\bibitem{Lazeyras:2015giz}
T. Lazeyras, M. Musso, and V. Desjacques, Phys. Rev. {\bf D93},  063007
  (2016).

\bibitem{moreno/giocoli/sheth:2009}
J. {Moreno}, C. {Giocoli}, and R.~K. {Sheth}, \mnras {\bf 397},  299  (2009).

\bibitem{robertson/kravtsov/etal:2009}
B.~E. Robertson, A.~V. Kravtsov, J. Tinker, and A.~R. Zentner, The
  Astrophysical Journal {\bf 696},  636  (2009).

\bibitem{eisenstein/seo/etal:2007}
D.~J. {Eisenstein}, H.-J. {Seo}, E. {Sirko}, and D.~N. {Spergel}, \apj {\bf
  664},  675  (2007).

\bibitem{kitaura/ensslin:2008}
F.~S. {Kitaura} and T.~A. {En{\ss}lin}, \mnras {\bf 389},  497  (2008).

\bibitem{kitaura/hess:2013}
F.-S. {Kitaura} and S. {He{\ss}}, \mnras {\bf 435},  L78  (2013).

\bibitem{jasche/wandelt:2013}
J. {Jasche} and B.~D. {Wandelt}, \mnras {\bf 432},  894  (2013).

\bibitem{schmittfull/feng/etal:2015}
M. {Schmittfull}, Y. {Feng}, F. {Beutler}, B. {Sherwin}, and M.~Y. {Chu}, \prd
  {\bf 92},  123522  (2015).

\bibitem{achitouv/blake:2015}
I. {Achitouv} and C. {Blake}, \prd {\bf 92},  083523  (2015).

\bibitem{keselman/nusser:2016}
A. {Keselman} and A. {Nusser}, ArXiv e-prints  (2016).

\bibitem{padmanabhan/xu/etal:2012}
N. {Padmanabhan}, X. {Xu}, D.~J. {Eisenstein}, R. {Scalzo}, A.~J. {Cuesta},
  K.~T. {Mehta}, and E. {Kazin}, \mnras {\bf 427},  2132  (2012).

\bibitem{kazin/koda/etal:2014}
E.~A. {Kazin}, J. {Koda}, C. {Blake}, N. {Padmanabhan}, S. {Brough}, M.
  {Colless}, C. {Contreras}, W. {Couch}, S. {Croom}, D.~J. {Croton}, T.~M.
  {Davis}, M.~J. {Drinkwater}, K. {Forster}, D. {Gilbank}, M. {Gladders}, K.
  {Glazebrook}, B. {Jelliffe}, R.~J. {Jurek}, I.-h. {Li}, B. {Madore}, D.~C.
  {Martin}, K. {Pimbblet}, G.~B. {Poole}, M. {Pracy}, R. {Sharp}, E.
  {Wisnioski}, D. {Woods}, T.~K. {Wyder}, and H.~K.~C. {Yee}, \mnras {\bf 441},
   3524  (2014).

\bibitem{burden/percival/etal:2014}
A. {Burden}, W.~J. {Percival}, M. {Manera}, A.~J. {Cuesta}, M. {Vargas Magana},
  and S. {Ho}, \mnras {\bf 445},  3152  (2014).

\bibitem{padmanabhan/white/cohn:2009}
N. {Padmanabhan}, M. {White}, and J.~D. {Cohn}, \prd {\bf 79},  063523  (2009).

\bibitem{vargas/ho/etal:2015}
M. {Vargas-Maga{\~n}a}, S. {Ho}, S. {Fromenteau}, and A.~J. {Cuesta}, ArXiv
  e-prints  (2015).

\bibitem{paranjape/sefusatti/etal:2013}
A. {Paranjape}, E. {Sefusatti}, K.~C. {Chan}, V. {Desjacques}, P. {Monaco}, and
  R.~K. {Sheth}, \mnras {\bf 436},  449  (2013).

\bibitem{lazeyras/wagner/etal:2016}
T. {Lazeyras}, C. {Wagner}, T. {Baldauf}, and F. {Schmidt}, \jcap {\bf 2},  018
   (2016).

\bibitem{zentner:2007}
A.~R. {Zentner}, International Journal of Modern Physics D {\bf 16},  763
  (2007).

\bibitem{dalal/white/etal:2008}
N. {Dalal}, M. {White}, J.~R. {Bond}, and A. {Shirokov}, \apj {\bf 687},  12
  (2008).

\bibitem{desjacques:2008}
V. {Desjacques}, \mnras {\bf 388},  638  (2008).

\bibitem{chan:2014}
K.~C. {Chan}, \prd {\bf 89},  083515  (2014).

\bibitem{shandarin/zeldovich:1989}
S.~F. {Shandarin} and Y.~B. {Zeldovich}, Reviews of Modern Physics {\bf 61},
  185  (1989).

\bibitem{biagetti/etal:2014}
M. {Biagetti}, K.~C. {Chan}, V. {Desjacques}, and A. {Paranjape}, \mnras {\bf
  441},  1457  (2014).

\bibitem{taylor/hamilton:1996}
A.~N. {Taylor} and A.~J.~S. {Hamilton}, \mnras {\bf 282},  767  (1996).

\end{thebibliography}
